\documentclass[aps,prl,reprint,groupedaddress]{revtex4-1}

\usepackage{graphicx}
\usepackage{color}
\usepackage{amsmath}

\begin{document}


\title{Local entropy and nonextensivity of networks ensembles}


\author{Qi Zhang}
\email[]{qi.zhang@just.edu.cn}
\affiliation{Theoretical Physics Research Center, School of Science, Jiangsu University of Science and Technology, Zhenjiang 212100, China}
\affiliation{Lorentz Institute for Theoretical Physics, Leiden University, PO Box 9504, 2300 RA Leiden, The Netherlands}

\author{Meizhu Li}
\email[]{Meizhu.li@ugent.be}
\affiliation{Department of Telecommunications and Information Processing, TELIN-GAIM, Ghent University, Gent, Belgium}


\date{\today}

\begin{abstract}
Nonextensivity is foreseeable in network ensembles, as heterogeneous interactions generally exist in complex networked systems that need to be described by network ensembles. But this nonextensivity has not been literatured proved yet. In this work, the existence of nonextensivity in the binary and weighted network ensembles is theoretically proved for the first time (both in the microcanonical and canonical ensemble) based on the finding of the local entropy's nonlinear change when new nodes are added to the network ensembles. This proof also shows that the existence of nonextensivity is the main difference between the network ensembles and other traditional models in statistical physics (Ising model). 
\end{abstract}

\pacs{}

\maketitle


\section{Introduction}
At the beginning of this century, network science as a new interdiscipline was built to describe those complex systems with networked structures, such as biological, social, and technological systems~\cite{newman2018networks,semrau2017dynamics}. It is a combination of mathematics, physics, computer science, and even social science~\cite{gosak2018network,kenett2015network,battiston2020networks,battiston2021physics,li2018evidential,rabinovich2020sequential,zhang2018measure}. Several decades of research on network science gives numerous new theoretical models and practical tools to explore those networked-complex systems.

Physicists play an important role in the research of network science. Many conceptions in physics have been transplanted to the research of network science to explore the physical characteristics of networked complex systems, e.g. phase transition~\cite{van2005phase,hanel2005phase,arenas2008synchronization,colomer2014double}, synchronization~\cite{arenas2008synchronization}, percolation~\cite{colomer2014double,jia2010post,cuquet2009entanglement}, and Boes-Einstein condensation~\cite{bianconi2001bose,crisanti2019condensation,garlaschelli2009generalized,zhang2022strong}. Simultaneously, the application of these fundamental conceptions from physics to network science also brings many new results to the research in traditional statistical physics, such as the breaking of ensemble equivalence with local constraints~\cite{cimini2019statistical,bianconi2013statistical,park2004statistical,dionigi2021spectral}, the finding of 'scale-free' property~\cite{strogatz2022fifty}. The combination of network science and physics, especially statistical physics, provides many new theoretical models, such as the network ensembles, which are the extension of statistical ensembles to network science. In network science, a network ensemble represents a set of networks that satisfies given structure properties, and its statistical property is the illustration of the macroscopic behaviors of those complex networked systems~\cite{jaynes1957information,bianconi2007entropy,cofre2018information,den2018breaking}. 

Traditionally, in statistical physics, systems with different constraints need to be described by the different statistical ensembles. For instance, systems with fixed total energy ${E}$ are described by the microcanonical ensemble, and systems with fixed temperature ${T}$ are described by the canonical ensemble. In network science, the constraints in each network are not the total energy or the fixed temperature. They are the number of links connected with each node or the number of triangles in the topological structure of each network~\cite{hollander2018breaking}. When the constraints of these networks have different properties, these networks also need to be described by different network ensembles. For example, in the binary networks represented by graph ensemble $G(N,L)$, the constraints are the total number of links $L$ and the total number of nodes $N$. When the total number of links $L$ for each binary network is fixed, the microcanonical binary network ensemble can be used to describe those networks. But when the constraint is the average number of links $\langle{L}\rangle$ in those networks, this set of networks needs to be described by the canonical binary network ensemble~\cite{bianconi2009entropy,zhang2022betweenness}. Same for the weighted networks, but the constraints are the total strength of links~\cite{newman2004analysis,serrano2006correlations,zhang2020ensemble}

Extensivity in natural systems can be explained as the existence of extensive properties in that systems~\cite{gibbs1902elementary,bialek2001complexity,binek2006nonextensivity,das2017extensivity}. It is typically represented by the linear increase of the structural characteristics following the growth of the system's size~\cite{dunning1985logical,mannaerts2014extensive,tsallis2022complex,tsekouras2004nonextensivity}. For instance, the entropy is an extensive property, the growth of the system's size will bring a linear increase in its entropy~\cite{gibbs1902elementary}. However, this extensivity is not always held, especially when the newly added units have multiple interactions with those existing units. Thus, checking the extensivity need to focus on the asymptotic behavior of each subsystem's entropy, i.e., the local entropy in the systems. 

Local entropy is the entropy of a subsystem, e.g., the entropy of the canonical ensemble is a local entropy for the whole system, which includes the subsystem that is described by the canonical ensemble and the heat bath~\cite{jaynes1957information,gibbs1902elementary}. When the subsystem is built based on each unit, the local entropy is a quantification of each unit's complexity. Then the behavior of the local entropy for each unit is the manifestation of the extensivity of the system. 
For instance, the Ising model with $n$ spins has $2^n$ configurations, and its microcanonical entropy is $S^{(n)}_{\textrm{mic}}=\ln 2^n$. When the number of spins in the Ising model increase to $n+1$, then the total number of configurations in it is $2^{n+1}$, and its new microcanonical entropy equals $S_{\textrm{mic}}^{(n+1)}=\ln 2^{n+1}$. Obviously, the growth of the Ising model's size causes a linear increase in its entropy, which means the Ising model is extensive. This extensivity can be represented as the local entropy of the $n$th spin $s_{n}=\ln 2$ equals to the growth on the entropy of the Ising model as $s_{n}=S_{\textrm{mic}}^{(n+1)}-S_{\textrm{mic}}^{(n)}$. This example inspired us how to check the extensivity in network ensembles, checking the relationship between the local entropy of the newly added unit in the network ensemble and the growth of the network ensemble's entropy~\cite{park2004statistical,bianconi2009entropy}. But to apply this method, we should define the units in network ensembles.

In networks, the nodes represent the units in the networked systems, and the links among those nodes represent the possible connection in those different units. Thus, it is reasonable to set the nodes with different constraints in the network as the units of the network ensembles, as a newly added node will bring at least one new link to the network when we set the network completely connected. However, when we set the number of links in the network as the size of the network ensemble, then a new add link in the network may not even bring a new add node to the network. Thus, using the number of nodes to represent the size of the network is reasonable in the structural complexity quantifying. Therefore, when the networks are under the same constraint (fixed $L$ or $\langle L\rangle$), the network ensemble with more nodes always has big entropy~\cite{bianconi2009entropy}. Then the existence of the extensivity in the network ensembles can be checked by the relationship between the local entropy of the newly added node and the growth of the entropy of the whole network ensemble, i.e., the relationship between the increase of the system's size and the entropy growth in the network ensemble. 

The research in this work will focus on the network ensembles' structural extensivity and local entropy. As the definition of the local entropy of the network ensembles is based on the properties of constraints in the network ensembles, we will introduce two local entropies of network ensembles (the microcanonical local entropy and the canonical local entropy of the network ensemble). And based on the different types of networks (binary and weighted networks), each local entropy will have two calculation forms. Thus, this work will introduce four different types of local entropies and try to discuss the extensivity of those network ensembles. And we will try to find extra information about the network ensembles' structural complexity from the behavior of the local entropy.

The rest of this paper is organized as follows: In section 2, two different ways of nodes' addition to the network ensemble are introduced.
Section 3 introduces the canonical local entropy for the binary (weighted) network ensembles and discusses its relationship with the network ensemble's nonextensivity. The definition of the microcanonical local entropy for the binary (weighted) network ensembles gives in section 4. It also includes the discussion of the nonextensivity in microcanonical network ensemble. Conclusions are in section 5. 

\section{Node's addition to network ensembles}
As mentioned above, the extensivity of network ensembles can be detected by the relationship between the increase of the network's size and its corresponding growth of entropy. Thus, how to add the new node to the network ensemble decides the structural complexity changing of the network ensemble. Therefore, to check the extensivity of network ensembles, the first problem that needs to be clarified in our work is how the extra nodes and their affiliate degree or total weights will be added to the network ensemble. 

\begin{figure}[htbp]
    \centering
        \includegraphics[width=8.6cm]{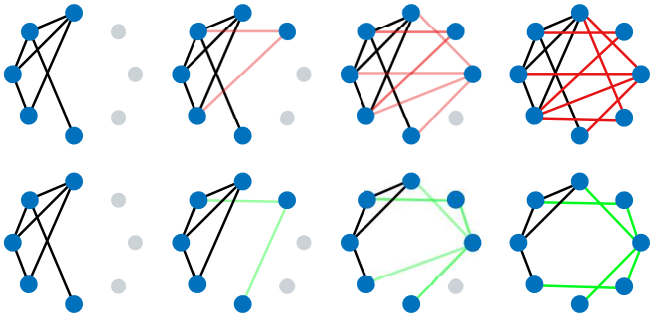}
        \caption{The process of the nodes adding to the binary network. First, we use the way that keeps the former topological structure and put new nodes and links in it. This process is shown in the top four subfigures. The original degree sequence for the network with five nodes is $\vec{k}=\{1,2,3,3,3\}$. When the node $6$ with degrees 2 adds to the network (the red links show in the top subfigures), the degree sequence becomes $\vec{k}^{}=\{1,3,3,4,3,2\}$. The degree of node $2$ and node $4$ changed. When the node $7$ with degrees $4$ is added to the network with six nodes, the degree sequences will change to $\vec{k}^{}=\{2,4,4,4,4,2,4\}$. When the node $8$ with degrees $2$ is added to the network with $7$ nodes, the degree sequences will change to $\vec{k}^{}=\{2,5,4,4,5,2,4,2\}$. 
        The second way of node's addition to the network ensemble is based on the condition that the degree sequence of the former nodes should keep the same. The bottom four subfigures show how the nodes and their affiliate links add to the network without changing the original degree sequences. The original degree sequences of the five nodes are $\vec{k'}^{}=\{1,2,3,3,3\}$. When the node $6$, $7$ and $8$ with degrees 2, 4 and 2 add to the network, the original degree sequences are still fixed as $\vec{k'}^{}=\{\textbf{1,2,3,3,3},2,4,2\}$.}
        \label{Two_ways_node_add}
\end{figure}

This section introduces two ways of the new node's addition to the network ensembles. The two ways of node addition details are shown in Fig.~\ref{Two_ways_node_add}.

The top four subfigures show how the nodes and their affiliate links (red links) add to the network under the condition that the former topological structure should keep in the new network. The bottom four subfigures show how the nodes and their affiliate links (green links) add to the network without changing the original degree sequences. The two different ways of nodes addition show two principles that are used to guide the network's growth: first, the old structure still exists in the network, but the main block has already changed (the first way); second, the old units in the network still retains its former local structural characteristics (from numeric value), but topological structure completely changed. 

The two ways of node addition will lead the network ensembles to have two different topological structural complexity. And the affiliated local entropy change is also different.

\section{Local entropy of canonical network ensemble}
The properties of the constraints in the thermodynamic systems decide which ensembles are used to describe those systems~\cite{squartini2015breaking}. When the constraints are hard, the systems need to be described by the microcanonical ensemble. When the constraints are soft, the systems must be described by the canonical ensemble. Thus, the local entropy of the network ensembles also has two different types, the local entropy of the microcanonical network ensemble and the local entropy of the canonical network ensembles. 

The canonical network ensemble is based on the assumption that the constraints are soft. The average value of each element in the degree sequences $C(\mathbf{G})=\{k_i\}$ is fixed and equal to the 'hard' constraints in the conjugate microcanonical network ensemble as $\langle k_i\rangle=k_i^*$. The probability distribution of the states in the canonical network ensemble should maximize the maximum entropy of it and realize the condition $\langle k_i\rangle=k_i^*$. Thus, probability of the network $\mathbf{G}$ in the canonical networks ensemble $\mathcal{G}_{\textrm{can}}(n,\vec{\theta^*})$ equals to 
\begin{equation}
    P_{\textrm{can}}(\mathbf{G})=\frac{e^{{H}(\mathbf{G},\vec{\theta^*})}}{Z(\vec{\theta^*})}.
\end{equation}
The constraints in the canonical network ensemble are the average value of the fixed degree of each node. The symbol $\vec{\theta^*}$ represents the maximum likelihood parameter, which is the solution of the realization of hard constraints and maximum Shannon entropy as
\begin{equation}
    \langle C(\mathbf{G})\rangle=\sum_{\mathbf{G}\in\mathcal{G}_{\textrm{can}}}C(\mathbf{G})P_{\textrm{can}}(\mathbf{G}).
\end{equation}
Symbol ${H}(\mathbf{G},\vec{\theta^*})$ represents the \textit{Hamilitonian} of each network in the ensemble~\cite{jaynes1957information}. It is a dot product between the constraints $C(\mathbf{G})$ and the likelihood parameter $\vec{\theta^*}$ as 
\begin{equation}
    {H}(\mathbf{G},\vec{\theta^*})=\sum_{i=1}^nk_i\theta^*_i.
\end{equation} 
The partition function $Z(\vec{\theta^*})$ of the canonical network ensemble with degree sequences $\{k^*_i\}$ and the maximum likelihood parameter $\vec{\theta^*}$ is 
\begin{equation}
    Z(\vec{\theta^*})=\sum_{\mathbf{G}\in\mathcal{G}_{\textrm{can}}}e^{-{H}(\mathbf{G},\vec{\theta^*})}.
\end{equation}
The Shannon entropy of the canonical ensemble with degree sequences $\{k_i\}$ can be calculated as 
\begin{equation}
    \begin{aligned}
         S_{\textrm{can}}&=-\sum_{\mathbf{G}\in\mathcal{G}_{\textrm{can}}}P_{\textrm{can}}(\mathbf{G})\ln P_{\textrm{can}}(\mathbf{G}).\\
                     &=\sum_{\mathbf{G}\in\mathcal{G}_{\textrm{can}}}P_{\textrm{can}}(\mathbf{G}){H}(\mathbf{G},\vec{\theta^*})+\ln Z(\vec{\theta^*})\\
                     &=\langle\vec{C}(\mathbf{G})\rangle\cdot \vec{\theta^*}+\ln Z(\vec{\theta^*})\\
                     &=\vec{C}^*\cdot\vec{\theta^*}+\ln Z(\vec{\theta^*}).
    \end{aligned}
\end{equation}
The calculations shown above prove that the construction of the constraints will affect the definition of the \textit{Hamilitonian} and the partition function $Z(\vec{\theta^*})$ of the canonical network ensemble. As each unit in the degree sequences $\{k^*_i\}$ is the sum of all the elements in each column or row of the adjacency matrix that affiliate with each network $\mathbf{G}$, the \textit{Hamilitonian} and partition function $Z(\vec{\theta^*})$ will be affected by the value of units in the adjacency matrix. Thus, the binary canonical network ensemble and the weighted canonical network ensemble will have different partition functions and different entropy. 

\subsection{Local entropy of binary canonical network ensemble}
In the binary network, the link between each pair of nodes is represented by the unit in the adjacency matrix with a value of 1 or 0. 
Each network that belongs to the canonical network ensemble is represented by the adjacency matrix $\mathbf{G}$, where each unit $g_{ij}\in[0,1]$, and the degree is the sum of all the elements in each column or row of the adjacency matrix as $k_i=\sum_{j=1}^{n}g_{ij}$. Then the partition function should equal to
\begin{equation}
    Z(\vec{\theta^*})=\sum_{\mathbf{G}\in\mathcal{G}_{\textrm{can}}}e^{-\sum_{i=1}^n\sum_{j=1}^ng_{ij}\theta^*_i}=\prod_{i=1}^{n}\prod_{j=1}^{n}(1+e^{-\theta^*_i}).
\end{equation}
The symbol $\theta^*_i$ is the maximum likelihood parameter affiliated with the $i$th unit in the constraints.

As the production based on $j$ can be simplified, the partition function of the binary network ensemble also can be simplified as 
\begin{equation}
    Z(\vec{\theta^*})=\prod_{i=1}^{n}(1+e^{-\theta^*_i})^n.
\end{equation}
It means the probability of each network $\mathbf{G}$ in the binary canonical network ensemble $\mathcal{G}_{\textrm{can}}$ can be calculated as the production of each link's probability as  
\begin{equation}
    P_{\textrm{can}}(\mathbf{G})=\prod_{i=1}^{n}\prod_{j=1}^{n}\frac{e^{-\theta^*_ig_{ij}}}{1+e^{-\theta^*_i}}.
\end{equation}
Probability of the link to connect node $i$ and node $j$ is $p_{ij}=\frac{e^{-\theta^*_i}}{1+e^{-\theta^*_i}}$, and simultaneously the probability of the link between node $i$ and node $j$ missed between node $i$ and node $j$ is equal to $p_{ij}=\frac{1}{1+e^{-\theta^*_i}}$. The two probabilities can be used to calculate the average value of the link weights between node $i$ and node $j$ in the binary canonical network ensemble. 

When we use the symbol $\langle g_{ij}\rangle$ to represent the average value of the weights of the link $g_{ij}$, it should equal the probability of each different kind of weights times weight as
\begin{equation}
    \begin{aligned}
    \langle g_{ij}\rangle&=\sum_{g_{ij}\in\mathbf{g}}p(g_{ij})g_{ij}\\
    &=0\times\frac{1}{1+e^{-\theta^*_i}}+1\times\frac{e^{-\theta^*_i}}{1+e^{\theta^*_i}}\\
    &=\frac{e^{-\theta^*_i}}{1+e^{\theta^*_i}},   
    \end{aligned}
\end{equation}
where $\mathbf{g}$ represents the set of the value that unit $g_{ij}$ can get. Then the calculation of the entropy can be formulated as the sum of each link's average value as 
\begin{equation}
    S_{\textrm{can}}=\sum_{i=1}^{n}\sum_{j=1}^{n}[\theta^*_i\langle g_{ij}\rangle+\ln(1+e^{-\theta^*_i})].
\end{equation}
Thus, if we set a local network that includes the possible links among node $i$ and other nodes, we can get a local entropy based on the canonical networks ensemble as 
\begin{equation}
    s^{\textrm{can}}_i=n\times [\theta^*_i\langle g_{ij}\rangle+\ln(1+e^{-\theta^*_i})],
\end{equation}
where $\langle g_{ij}\rangle$ is the average value of the weights of the link $g_{ij}$. According to the definition in the canonical network ensemble, the sum of the average value of the weights of the possible links between node $i$ is the average value of the degree $k_i^*$, we can get the relationship between the parameter $\theta^*_i$ and affiliate degree $k^*_i$ of node $i$ as 
\begin{equation}
    k^*_i=n\times\langle g_{ij}\rangle=n\times\frac{e^{-\theta^*_i}}{1+e^{\theta^*_i}}, e^{-\theta^*_i}=\frac{k^*_i}{n-k^*_i}.
\end{equation}

In binary networks, the weight of the link in the network only has two choices, value $1$ or value $0$. The probability $p_{ij}$ is also decided by the value of weights, the likelihood parameter, and the possible combinations. 
Then we can find the local canonical entropy of the networks with degree sequences $\{k^*_i\}$ is 
\begin{equation}\label{local_can_theta}
    s^{\textrm{can}}_i=n\times [\theta^*_i\frac{e^{-\theta^*_i}}{1+e^{\theta^*_i}}+\ln(1+e^{-\theta^*_i})].
\end{equation}
The value of $\theta^*_i$ can be calculated from the relationship between the average value of $g_{ij}$ and the fixed degree $k_i^*$. The hard constraint in the microcanonical networks is equal to the average value of the degree in the network in the canonical ensemble. Thus, according to the relationship between the hard constraints and the average value of degrees in the canonical ensemble, $k_i^*=\langle k_i\rangle$, the value of $\theta^*_i$ is equal to 
\begin{equation}\label{theta_i}
    \theta^*_i=-\ln\frac{n-k_i^*}{k_i^*}.
\end{equation}
When put the Eq.\eqref{theta_i} into Eq.\eqref{local_can_theta}, the value of the local entropy of the canonical networks ensemble is a function of degree $k^*_i$, which equals to 
\begin{equation}\label{s_can_k_i}
    \begin{aligned}
        s^{\textrm{can}}_i&={k^*_i}\ln\frac{n-k^*_i}{k^*_i}+n\ln\frac{n}{n-k^*_i}\\
        &=\ln\frac{n^n}{{k^*_i}^{k^*_i}(n-k^*_i)^{n-k^*_i}}
    \end{aligned}
\end{equation}
The calculation of the canonical entropy also can be reformulated as 
\begin{equation}
    S_{\textrm{can}}=\sum_{i=1}^ns^{\textrm{can}}_i=\sum_{i=1}^n[{(n-k^*_i)}\ln\frac{n-k^*_i}{k^*_i}+n\ln\frac{n}{k^*_i}].
\end{equation}
This result means the canonical entropy of the network ensemble with degree sequence $\{k^*_i\}$ is extensive when the size of the networks is fixed. And obviously, its value will increase following the growth of the network's size. 

\subsubsection{The nonextensivity and local entropy in binary canonical network ensemble}
As shown above, the node addition to the canonical network ensemble has two different ways. The discussion of the extensivity in the binary canonical network ensemble also needs to be divided into two parts. 

The first part of the discussion is based on the first way of node addition, which will change the original degree sequences of the binary canonical network ensemble. Thus, when the new node $n+1$ with degree $k^*_{n+1}$ add to the network with $n$ nodes, there will be $k^*_{n+1}$ nodes' degree increase one. This change will cause an alteration in the value of the probability for the links to appear among different pairs of nodes. The change in the degree sequences does not change the form of the partition function, but the value of the conjugate maximum likelihood parameter of each unit in the constraint will change. Thus, when the network has $n+1$ nodes, the probability of each state $\mathbf{G}$ equals 
\begin{equation}
    {P'}_{\textrm{can}}(\mathbf{G})=\prod_{i=1}^{n+1}\prod_{j=1}^{n+1}\frac{e^{-{\theta'}^*_{i}g_{ij}}}{1+e^{-{\theta'}^*_{i}}}.
\end{equation}

In the probability ${P'}_{\textrm{can}}(\mathbf{G})$, there are $n+1$ maximum likelihood parameters. When all the nodes connect with the newly added node with label $n+1$ are included in the set $\mathcal{E}$, the value of the node $t$'s maximum likelihood parameter $\theta'^*_t$ can be calculated from the new degree $k^*_t=k^*_i+1$ (to simplify the calculation we still use $k^*_i+1$ in the calculation, and here the symbol $k^*_i$ represents the former degree of the nodes that is belonged to the set $\mathcal{E}$) as
\begin{equation}
    \frac{e^{-{\theta'}^*_{t}}}{1+e^{-{\theta'}^*_{t}}}=\frac{k^*_i+1}{n+1}, e^{-{\theta'}^*_{t}}=\frac{k^*_i+1}{n-k^*_i}
\end{equation}
Then the local entropy of node $t$ will change as the maximum likelihood parameter has already changed. The new local entropy of node $t$ ${s'}_{t}^{\textrm{can}}$ is 
\begin{equation}
    \begin{aligned}
    {s'}_{t}^{\textrm{can}}&=(k^*_i+1)\ln\frac{n-k^*_i}{k^*_i+1}+(n+1)\ln\frac{n+1}{n-k^*_i}\\
    &= \ln\frac{(n+1)^{n+1}}{(k^*_i+1)^{k^*_i+1}(n-k^*_i)^{n-k^*_i}}.
    \end{aligned}
\end{equation}
There is $k^*_{n+1}$ nodes have a new add link that is connected with the node $n+1$. The sum of the local entropy of the nodes in this set $\mathcal{E}$ should equal to
\begin{equation}
    \begin{aligned}
    {S'}(\mathcal{E})&=\sum_{t\in\mathcal{E}}{s'}_{t}^{\textrm{can}}\\
    &=\sum_{t\in\mathcal{E}}\ln\frac{(n+1)^{n+1}}{(k^*_i+1)^{k^*_t+1}(n-k^*_i)^{n-k^*_i}}.        
    \end{aligned}
\end{equation}

There still have other $n-k^*_i$ nodes in the network ensemble whose degree does not change. But their local entropy also changes when the new node $n+1$ is added to the binary network ensemble. These nodes belong to set $\mathcal{O}$, and the maximum likelihood parameter ${\theta'}_h^*$ of node $h$ in set $\mathcal{O}$ equals to 
\begin{equation}
    \frac{e^{-{\theta'}_h^*}}{1+e^{-{\theta'}_h^*}}=\frac{k^*_i}{n+1},e^{-{\theta'}_h^*}=\frac{k^*_i}{n+1-k^*_i}.
\end{equation}
Here, the calculation of ${\theta'}_h^*$ is still based on $k^*_i$, and the $k^*_i$ represents the degree of every possible node that is belonged to the set $\mathcal{O}$.
According to the change in the parameter ${\theta'}_t^*$, the local entropy of the node in the set $\mathcal{O}$ should equal to 
\begin{equation}\label{s_h_o}
    \begin{aligned}
    {s'}^{\textrm{can}}_h&=k^*_i\ln\frac{n+1-k^*_i}{k^*_i}+(n+1)\ln\frac{n+1}{n+1-k^*_i}\\
    &=\ln\frac{(n+1)^{n+1}}{{k^*_i}^{k^*_i}(n+1-k^*_i)^{n+1-k^*_i}}
    \end{aligned}
\end{equation}
The sum of the local entropy of all the nodes in set $\mathcal{O}$ is 
\begin{equation}
    \begin{aligned}
    {S'}(\mathcal{O})&=\sum_{h\in\mathcal{O}}{s'}^{\textrm{can}}_h\\
    &=\sum_{h\in\mathcal{O}}\ln\frac{(n+1)^{n+1}}{{k^*_k}^{k^*_i}(n+1-k^*_i)^{n+1-k^*_i}}. 
    \end{aligned}
\end{equation}
It is clear that the local entropy of each node in the binary network ensemble is increased, and adding one node to the network will bring a change in local entropy for every node. The nodes in the set $\mathcal{E}$ increase more than the nodes in set $\mathcal{O}$.

The Shannon entropy of the binary network ensemble with a newly added node with degree $k^*_{n+1}$ equals to the sum of the entropy ${S'}(\mathcal{O})$ and ${S'}(\mathcal{E})$ as 
\begin{equation}
    S_{\textrm{can}}'={S'}(\mathcal{O})+{S'}(\mathcal{E}).       
\end{equation}

The increase of the local entropy of nodes in the original network ensemble also can be divided into two sets. In the set $\mathcal{E}$, the grows of the node's local entropy can be calculated as 
\begin{equation}
    \Delta{s'}_t^{\textrm{can}}={s'}_t^{\textrm{can}}-s^{\textrm{can}}_t=\ln\frac{(n+1)^{n+1}{k^*_i}^{k^*_i}}{n^n(k^*_i+1)^{k^*_i+1}}.
\end{equation}

In the set $\mathcal{O}$, the growth of each node's entropy is also decided by the former degree of that node, plus some correction part as 
\begin{equation}
    \Delta{s'}_h^{\textrm{can}}={s'}_h^{\textrm{can}}-s^{\textrm{can}}_h=\ln\frac{(n+1)^{n+1}{(n-k^*_i)}^{n-k^*_i}}{n^n(n-k^*_i+1)^{n-k^*_i+1}}.
\end{equation}

\begin{figure}[htbp]
    \centering
        \includegraphics[width=4.3cm]{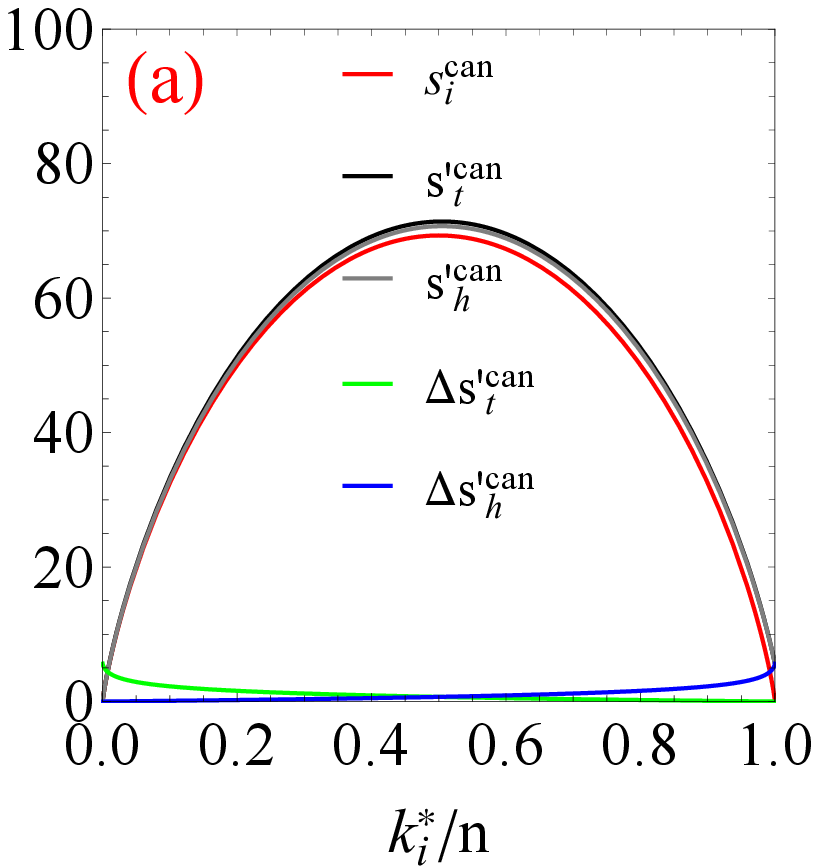}
        \includegraphics[width=4.15cm]{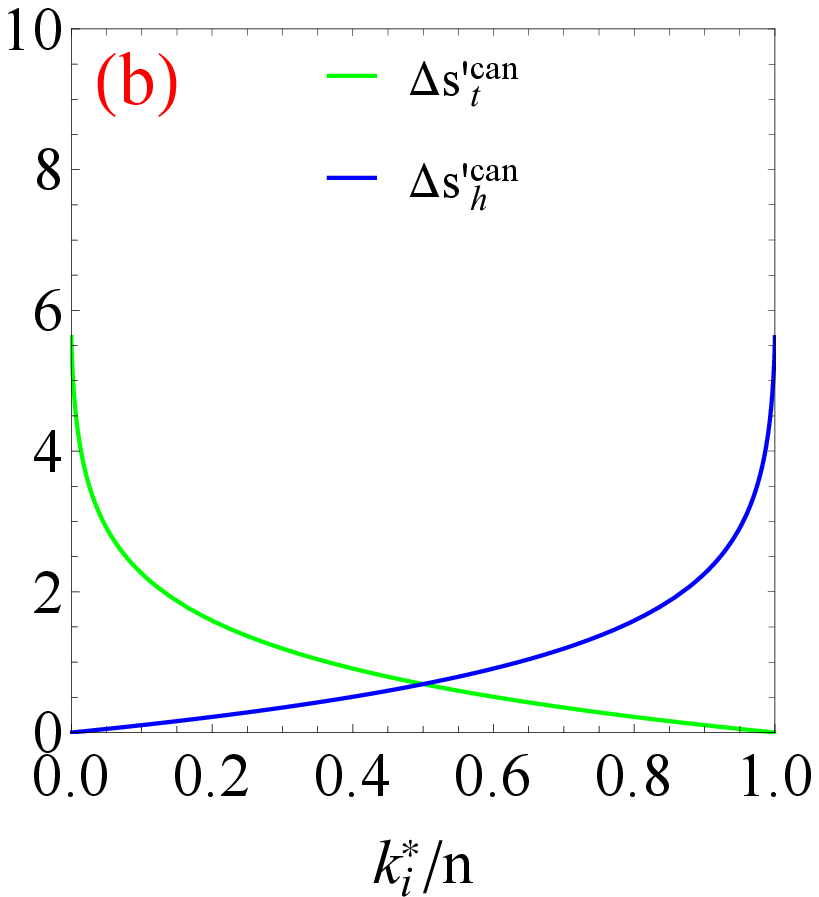}
        \caption{The asymptotic behavior of local entropy in the binary network ensemble (The original network has 100 nodes). Subfigure (a) shows the asymptotic behavior of the local entropy in the binary canonical network ensemble. We can find that the local entropy of node $i$ does not have a big difference, and they are symmetry with the ratio between $k^*_i$ and $n$. when the value of ${k^*_i}/{n}$ is equal to $0.5$, the local entropies have the biggest value (both for the $s^{\textrm{can}}_{i}$, ${s'}^{\textrm{can}}_{t}$ and ${s'}^{\textrm{can}}_{h}$). The node's local entropy growth under the first way of node addition has two different paths. When the nodes are included in the set $\mathcal{O}$, the bigger the original degree of the node, the bigger the value of the local entropy growth. However, in the set of $\mathcal{E}$, the bigger the original degree of the node, the smaller the increase of local entropy (shown in subfigure (b)).}
        \label{First_Binary_local_entropy_can}
\end{figure}

The asymptotic behavior of the local entropy and its growth in the binary network ensemble is still constrained by the total number of nodes in the network ensemble. These details are shown in Fig.\ref{First_Binary_local_entropy_can}.

The asymptotic behavior of $\Delta{s'}^{\textrm{can}}_t$ and $\Delta{s'}^{\textrm{can}}_h$ have a big difference. If the nodes belong to the set $\mathcal{E}$, which nodes are directly connected with the new add node, the bigger the former degree, the smaller the growth of the local entropy. As the value of the $\Delta{s'}^{\textrm{can}}_t$ is decided by ${{k^*_i}^{k^*_i}}/{(k^*_i+1)^{k^*_i+1}}$, the bigger the value of $k^*_i$ the smaller the value of ${{k^*_i}^{k^*_i}}/{(k^*_i+1)^{k^*_i+1}}$. However, the subleading part in $\Delta{s'}^{\textrm{can}}_h$ is ${(n-k^*_i)^{n-k^*_i}}/{(n-k^*_i+1)^{n-k^*_i+1}}$ when the number of nodes in this ensemble is fixed. The value of the subleading part in $\Delta{s'}^{\textrm{can}}_h$ is decided by the degree of the node $k^*_i$. The increase of the value of $k^*_i$ will cause the growth of the local entropy. The mathematical reasons are already shown above, but the physical means of the asymptotic behavior are still missing. 

The second way of node addition is different from the first one. This process will keep the degree of the former nodes in the network ensemble fixed. It means the change of each node's local entropy is due to the increase of the possible nodes that can be connected. 
When there is a new node with degree $k_{n+1}^*$ added to the network ensemble, the probability of each network satisfying the constraints will equal to 
\begin{equation}
    \tilde{P}_{\textrm{can}}(\mathbf{G})=\frac{e^{-\tilde{\theta}^*_{n+1}g_{ij}}}{1+e^{-\tilde{\theta}^*_{n+1}}}\prod_{i=1}^{n}\prod_{j=1}^{n}\frac{e^{-\tilde{\theta}^*_ig_{ij}}}{1+e^{-\tilde{\theta}^*_i}},
\end{equation}
where $\tilde{\theta}^*_{n+1}$ is the corresponding maximum likelihood parameter of the new added node with degree $k_{n+1}^*$. When the new degree sequence with $n+1$ units is represented by $\vec{C}(\mathbf{G})$, the Shannon entropy of the binary canonical network ensemble with $n+1$ nodes can be reformulated as  
\begin{equation}
    \begin{aligned}
        \tilde{S}_{\textrm{can}}&=-\sum_{\mathbf{G}\in\mathcal{G}_{\textrm{can}}}\tilde{P}_{\textrm{can}}(\mathbf{G})\ln \tilde{P}_{\textrm{can}}(\mathbf{G})\\
        &=\langle \vec{C}(\mathbf{G})\rangle\cdot\vec{\tilde{\theta}}^*+\ln Z(\vec{\tilde{\theta}}^*)\\
        &=\sum_{i=1}^{n+1}\sum_{j=1}^{n+1}[\langle \tilde{g}_{ij}\rangle\vec{\tilde{\theta}}^*+\ln(1+e^{-{\tilde{\theta}}^*_i})]\\
        &=(n+1)\times\sum_{i=1}^{n+1}[\langle \tilde{g}_{ij}\rangle\vec{\tilde{\theta}}^*+\ln(1+e^{-{\tilde{\theta}}^*_i})]\\
    \end{aligned}
\end{equation}
Symbol $\langle\tilde{g}_{ij}\rangle$ represents the average value of the weight between each pair of nodes. It will also change in the new network ensemble with $n+1$ nodes. And this change will cause the growth of the local entropy for the nodes in the new network ensemble. 

The value of $\langle\tilde{g}_{ij}\rangle$ equals to  
\begin{equation}\label{k_can_theta}
    \langle\tilde{g}_{ij}\rangle=\frac{k^*_i}{n+1}=\frac{e^{-{\tilde{\theta}}^*_i}}{1+e^{-{\tilde{\theta}}^*_i}}, e^{-{\tilde{\theta}}^*_i}=\frac{k^*_i}{n-k^*_i+1}
\end{equation} 
The local entropy of node $i$ is the sum of all the possible links among $n+1$ nodes in the new network ensemble, including node $i$ itself. The value of the local entropy $\tilde{s}^{\textrm{can}}_{i}$ is decided by $\tilde{\theta}^*_i$, which is a function of $k^*_i$ shown in Eq.~\eqref{k_can_theta}. The details of the definition of $\tilde{s}^{\textrm{can}}_{i}$ is 
\begin{equation}
    \begin{aligned}
        \tilde{s}^{\textrm{can}}_{i}&=(n+1)\times[\tilde{\theta}^*_i\frac{e^{-\tilde{\theta}^*_i}}{1+e^{-\tilde{\theta}^*_i}}+\ln(1+e^{-\tilde{\theta}^*_i})]\\
        &=(n+1)\times[\frac{k^*_i}{n+1}\ln\frac{n+1-k^*_i}{k^*_i}+\ln\frac{n+1}{n+1-k^*_i}]\\
        &=k^*_i\ln\frac{n+1-k^*_i}{k^*_i}+(n+1)\ln\frac{n+1}{n+1-k^*_i}\\
        &=\ln\frac{(n+1)^{n+1}}{{k^*_i}^{k^*_i}(n+1-k^*_i)^{n+1-k^*_i}}.
    \end{aligned}
\end{equation}
The new local entropy $\tilde{s}^{\textrm{can}}_{i}$ is equal to the local entropy ${s'}^{\textrm{can}}_h$ of the nodes in the set $\mathcal{O}$, which is definied in the first way of node addition show above that is introduced in Eq.~\eqref{s_h_o}. The asymptotic behavior of $\tilde{s}^{\textrm{can}}_{i}$ also can be found in Fig.~\ref{First_Binary_local_entropy_can}.

Then we can get the Shannon entropy of the new binary network ensemble with $n+1$ nodes as a function of $\{k^*_i\}$ and $n$  
\begin{equation}
    \tilde{S}_{\textrm{can}}=(n+1)\times\sum_{i=1}^{n+1}\ln\frac{(n+1)^{n+1}}{{k^*_i}^{k^*_i}(n+1-k^*_i)^{n+1-k^*_i}}.
\end{equation}

According to the definition of Shannon entropy, the change of the two network ensembles' Shannon entropy comes from the difference in each node's local entropy caused by the newly added node plus the local entropy of the newly added node. Thus, the difference between the Shannon entropy of the original network ensemble and the network ensemble with $n+1$ nodes shows the increase in the structural complexity that is caused by the newly added node. And this difference is equal to 
\begin{equation}
    \begin{aligned}
    \Delta S_{\textrm{can}}&=\tilde{S}_{\textrm{can}}-S_{\textrm{can}}\\  
    &=\tilde{s}^{\textrm{can}}_{n+1}+\sum_{i=1}^n\Delta \tilde{s}^{\textrm{can}}_{i}   
    \end{aligned}
\end{equation}
The $\tilde{s}^{\textrm{can}}_{n+1}$ is the local entropy of the newly added node with label $n+1$. Symbol $\Delta \tilde{s}^{\textrm{can}}_{i}$ represents the difference of the local entropy of node $i$ in the original network ensemble and the new network ensemble with $n+1$ nodes,
\begin{equation}
    \begin{aligned}
    \Delta \tilde{s}^{\textrm{can}}_{i}&=\tilde{s}^{\textrm{can}}_{i}-s^{\textrm{can}}_{i}\\
    &=\ln\frac{(n+1)^{n+1}{(n-k^*_i)}^{n-k^*_i}}{n^n(n+1-k^*_i)^{n+1-k^*_i}}\\
    &>0
    \end{aligned}
\end{equation}
The value of the growth of local entropy for the node in the original network is also equal to the increase of the local entropy for the nodes in the set $\mathcal{O}$ in the first way of node addition. And its asymptotic behavior is the same with $\Delta{s'}_h^{\textrm{can}}$, which is also shown in Fig.~\ref{First_Binary_local_entropy_can}.

Thus, the entropy of the binary canonical network ensemble is not extensive, as the increase of the system's size not only brings the entropy increase affiliated with the newly added nodes but also affects the local entropy of other existing nodes in the ensemble. The more nodes added to the network ensemble, the bigger the difference in their local entropy. Thus, the binary canonical network ensemble is nonextensive from the behavior of the local entropy. 

\subsection{Local entropy of weighted canonical network ensemble}
The weighted networks also play an important role in network science. The weighted links among different pairs of nodes allowed those networks that belong to the same ensemble have a heterogeneous structure. When the weight of each link in the network is equal to 1, this weighted network is the binary network, i.e., the binary network is a particular case of the weighted networks. Therefore, the states' distribution in the weighted network is a generalization of the binary one. Thus, checking if the nonextensivity already found in the binary network ensemble still existed in the weighted network ensemble is significant. Therefore, in this part, we will discuss how the structural complexity changes in the weighted networks when a new node is added to the network.

The \textit{Hamilitonian} in the weighted network with nodes' weights sequences $\vec{C}(\mathbf{G})=\{w^*_i\}$ is equal to the dot product of the nodes' total weights sequences and the conjugate maximum likelihood parameter. 

The total weights sequences still equal the sum of the weights of all the links that connect with node $i$ as $w_i=\sum_{j=1}^na_{ij}$, where symbol $a_{ij}$ represents the weights of the link between node $i$ and node $j$. The definition of the \textit{Hamilitonian} $H(\mathbf{G},\vec{\theta^*})$ is
\begin{equation}
    \begin{aligned}
    H(\mathbf{G},\vec{\theta^*})&=\vec{C}(\mathbf{G})\cdot\vec{\theta^*}=\sum_{i=1}^n\sum_{j=1}^na_{ij}\theta^*_i.
    \end{aligned}
\end{equation}
Each configuration $\mathbf{G}$ in the weighted canonical network ensemble $\mathcal{G}_{\textrm{can}}$ has its own $H(\mathbf{G},\vec{\theta^*})$. Then the partition function of the weighted canonical network ensemble 
\begin{equation}\label{parti_can_weigh}
    \begin{aligned}
        Z(\vec{\theta}^*)&=\sum_{\mathbf{G}\in\mathcal{G}_{\textrm{can}}}e^{-H(\mathbf{G},\vec{\theta}^*)}\\
        &=\prod_{i=1}^n\prod_{i=1}^n\sum_{a_{ij}\in{\omega}}e^{-a_{ij}\theta^*_i}\\       
    \end{aligned}
\end{equation}
is a function of parameter vector $\vec{\theta^*}$.

The symbol $\omega$ represents the link in the network that can choose the set of weights, and it is the natural integer (including 0). Then the partition function is equal to 
\begin{equation}
    \begin{aligned}
     Z(\vec{\theta}^*)&=\prod_{i=1}^n\prod_{j=1}^n\sum_{a_{ij}=0}^{\infty}e^{-a_{ij}\theta^*_i}\\
     &=\prod_{i=1}^n\prod_{j=1}^n\frac{1}{1-e^{-\theta^*_i}}.
    \end{aligned}
\end{equation}

The maximum value of the weights for the links in the weighted network is $\infty$, so the partition function is the limit of the function shown in Eq.\eqref{parti_can_weigh} when the value $a_{ij}$ grows from $0$ to $\infty$.
The definition of the \textit{Hamilitonian} and partition function allowed us to calculate the probability of network $\mathbf{G}$ in the network ensemble as
\begin{equation}
    P_{\textrm{can}}(\mathbf{G})=\frac{e^{-H(\mathbf{G},\vec{\theta^*})}}{Z(\vec{\theta}^*)}=\prod_{i=1}^n\prod_{j=1}^ne^{-\theta^*_ia_{ij}}{(1-e^{-\theta^*_i})}.
\end{equation}
This equation also shows the probability $p_{ij}$ of each link in the weighted network ensemble having different weights. It equals to 
\begin{equation}
    p_{\textrm{can}}(a_{ij})=e^{-\theta^*_ia_{ij}}{(1-e^{-\theta^*_i})}.
\end{equation}
The average value of the weight for the link between node $i$ and node $j$ in this weighted network is decided by the possible value of $a_{ij}$ and the probability of it appearing in the network $p_{\textrm{can}}(a_{ij})$. It is equal to
\begin{equation}
    \begin{aligned}
    \langle a_{ij}\rangle&=\sum_{a_{ij}=0}^\infty a_{ij}\times p_{\textrm{can}}(a_{ij})\\
    &=\sum_{a_{ij}=0}^\infty a_{ij}\times e^{-\theta^*_ia_{ij}}{(1-e^{-\theta^*_i})}\\
    &=\frac{e^{-\theta^*_i}}{1-e^{-\theta^*_i}}.
    \end{aligned}
\end{equation}
Because in the weighted network, the average value of the link $a_{ij}$'s weight still equals $\langle a_{ij}\rangle=\frac{w^*_i}{n}$, the value of the parameter $\theta^*_i$ is 
\begin{equation}
    e^{-\theta^*_i}=\frac{w^*_i}{n+w^*_i}.
\end{equation}

Then we can get the value of the Shannon entropy for the weighted canonical network ensemble with the weights sequences $\{w^*_i\}$ as 
\begin{equation}
    \begin{aligned}
        S_{\textrm{can}}&=\sum_{i=1}^n\sum_{j=1}^n\langle a_{ij}\rangle\cdot\theta^*_i+\ln Z(\vec{\theta^*})\\
        &=\sum_{i=1}^n\sum_{j=1}^n[\theta^*_i\frac{e^{-\theta^*}}{1-e^{-\theta^*_i}}+\ln\frac{1}{1-e^{-\theta^*_i}}]\\
        &=\sum_{i=1}^n[w^*_i\ln\frac{n+w^*_i}{w^*_i}+n\ln\frac{n+w^*_i}{n}]\\
        &=\sum_{i=1}^n\ln\frac{(n+w^*_i)^{n+w^*_i}}{{w^*_i}^{w^*_i}n^n}
    \end{aligned}
\end{equation}

As there are $n$ nodes in the weighted network ensemble, the local entropy of the node $i$ in the weighted canonical network ensemble is already shown above. Its value is based on the number of nodes in the weighted network and the total strength, which shows as $w^*_i$ in the equation as
\begin{equation}
    s^{\textrm{can}}_i=\ln\frac{(n+w^*_i)^{n+w^*_i}}{{w^*_i}^{w^*_i}n^n}.
\end{equation}
The local entropy of each node in the weighted canonical network ensemble is constrained by the total number of nodes in the networks and their total weights. Thus, the change in the network's size and the total consequences will cause a change in its local entropy.

\subsubsection{Nonextensivity and local entropy in weighted canonical network ensemble}
The checking of the extensivity in the weighted canonical network ensemble still needs us to find what will happen to the Shannon entropy and local entropy in the network ensemble when a new node with total strength $w^*_{n+1}$ is added to the network ensemble. 
Because the fundamental principle of the weights allocation is not changed, we can still get the partition function of the new network ensemble as 
\begin{equation}
    Z'(\vec{{\theta'}^*})=\prod_{i=1}^{n+1}\prod_{j=1}^{n+1}\frac{1}{1-e^{-\theta'^*_i}},
\end{equation}
which has $(n+1)\times(n+1)$ compoments. 

The probability of the weighted network to have state $\mathbf{G}$ with $n+1$ nodes in this ensemble is 
\begin{equation}
    P_{\textrm{can}}(\mathbf{G})=\prod_{i=1}^{n+1}\prod_{j=1}^{n+1}e^{-\theta'^*_ia'_{ij}}(1-e^{-\theta'^*_i}).
\end{equation}
Both under the two ways of node addition, the calculation of the partition function and the probability $P_{\textrm{can}}(\mathbf{G})$ have the same form. However, different ways of node addition will change the value of the weights' sequences $\vec{C}(\mathbf{G})=\{w'^*_{i}\}$ of the network ensemble. Thus, the average value of $a'_{ij}$ and the maximum likelihood parameter $\theta'^*_i$ have different values when we use different ways to add nodes to the network ensemble. 

The addition of new nodes to the weighted network ensemble still has two ways, like what is shown in Fig.~\ref{Two_ways_node_add}, the first way with the fixed-original structure and the second way with the fixed total weights for the original nodes. Thus, the discussion of the nonextensivity of the weighted canonical network ensemble still needs to be divided into two parts, each part affiliated with one way of the node's addition. 

The first part is based on the first way of node addition. When the new node and its affiliate total weights add to the network ensemble, the total weights of nodes connected with the newly added node will change. Thus, all the $n+1$ nodes can be divided into two sets: the set $\mathcal{E}$, which includes the nodes whose total weights are increased $\Delta w^*_i$, and the set $\mathcal{O}$, which includes nodes whose total weights are fixed.

Then we can find that the average value of the weights $a'_{ij}$ of the links between node $i$ (in set $\mathcal{E}$) and node $j$ is equal to 
\begin{equation}\label{weighted_average_can_theta}
    \langle a'^{(\mathcal{E})}_{ij}\rangle=\frac{e^{-\theta'^*_t}}{1-e^{-\theta'^*_t}}=\frac{w^*_i+\Delta w^*_i}{n+1},
\end{equation}
where $\theta'^*_t$ represents the maximum likelihood parameter affiliates with the total weights of node $i$ in the set $\mathcal{E}$ (which is represented by $\theta^*_i$ in the original network ensemble).

When we use $w'^*_i=w^*_i+\Delta w^*_i$ represents the new total weights of node $i$, which belongs to the set $\mathcal{E}$.
Results in Eq.\eqref{weighted_average_can_theta} can be reformulated as the function of $w'^*_i$ to calculate the value of the maximum likelihood parameter $\theta'^*_i$ in the new weighted network ensemble as
\begin{equation}
    e^{-\theta'^*_t}=\frac{w'^*_i}{n+1+w'^*_i}.
\end{equation}
The local entropy of the nodes in the weighted network ensemble is still based on the value of the parameter $\theta'^*_i$ as 
\begin{equation}
    \begin{aligned}
    s'^{\textrm{can}}_t&=(n+1)\times[\theta'^*_t\frac{e^{-\theta'^*_t}}{1-e^{-\theta'^*_t}}+\ln\frac{1}{1-e^{-\theta'^*_t}}]\\
        &=\ln\frac{(n+1+w'^*_i)^{n+1+w'^*_i}}{{w'^*_i}^{w'^*_i}(n+1)^{n+1}}.        
    \end{aligned}
\end{equation}
Here, we use the symbol $s'^{\textrm{can}}_t$ to represent the local entropy of the node $i$ in the set $\mathcal{E}$ of the canonical weighted network with $n+1$ nodes when the first way adds the new node.

The new total weights $w'^*_i$ is replaced by $w^*_i+\Delta w^*_i$, the value of the local entropy of node $i$ in the set $\mathcal{E}$ is 
\begin{equation}
    s'^{\textrm{can}}_t=\ln\frac{(n+1+w^*_i+\Delta w^*_i)^{n+1+w^*_i+\Delta w^*_i}}{{(w^*_i+\Delta w^*_i)}^{w^*_i+\Delta w^*_i}(n+1)^{n+1}}
\end{equation}
Therefore, we can find that the local entropy of the nodes in the set $\mathcal{E}$ is decided by the value of its former total weights $w^*_i$, the growth of its total weights $\Delta w^*_i$ and the total number of nodes in the ensemble $n+1$.

For the nodes in the set $\mathcal{O}$, whose total weights keep fixed when the new nodes add to the network ensemble, the average value of its 
total weights will be changed as the added node will affect the network's size. Thus, the average value equals to
\begin{equation}\label{average_value_a_o}
    \langle a'^{(\mathcal{O})}_{ij}\rangle=\frac{e^{-\theta'^*_h}}{1-e^{-\theta'^*_h}}=\frac{w^*_i}{n+1},
\end{equation}
where the symbol $\theta'^*_h$ represents the maximum likelihood parameter affiliated with the total weights of node $i$ in the set $\mathcal{O}$, which is represented by $\theta^*_i$ in the old network ensemble with $n$ nodes. Here, the node $i$ does not represent the specific node with the label $i$. It is a general reference to the node in the set.

The change in the average value of total weights is caused by the difference in the denominator $n+1$. 
This difference will affect the change in the local entropy of the node in the set $\mathcal{O}$. When we use the symbol $s'^{\textrm{can}}_h$ to represent the local entropy of the node $i$ in the set $\mathcal{O}$ of the new network, it equals to 
\begin{equation}
    \begin{aligned}
    s'^{\textrm{can}}_h&=(n+1)\times[\theta'^*_i\frac{e^{-\theta'^*_h}}{1-e^{-\theta'^*_h}}+\ln\frac{1}{1-e^{-\theta'^*_h}}]\\
    &=\ln\frac{(n+1+w^*_i)^{n+1+w^*_i}}{{w^*_i}^{w^*_i}(n+1)^{n+1}}.
    \end{aligned}
\end{equation}

The new Shannon entropy $S'_{\textrm{can}}$ of the weighted networks is equal to the sum of the local entropy of nodes in the two sets ($\mathcal{E}$ and $\mathcal{O}$, the node $n+1$ belongs to the set $\mathcal{O}$). When the total number of nodes in set $\mathcal{E}$ is $l$, then the number of nodes in the set $\mathcal{O}$ is $n+1-l$. Then the Shannon entropy of the new network ensemble with $n+1$ is equal to 
\begin{equation}
    \begin{aligned}
        S'_{\textrm{can}}=&\sum_{i=1}^{n+1}\sum_{j=1}^{n+1}[\theta'^*_i\frac{e^{-\theta'^*_i}}{1-e^{-\theta'^*_i}}+\ln\frac{1}{1-e^{-\theta'^*_i}}]\\
        =&\sum_{t=1}^l s'^{\textrm{can}}_t+\sum_{h=1}^{n+1-l}s'^{\textrm{can}}_h\\
    \end{aligned}
\end{equation}

As the increase of each node's local entropy is not homogenous, thus the growth of the local entropy for the nodes in different sets is also different. The increase of the local entropy for the nodes in the set $\mathcal{E}$ is 
\begin{equation}
    \begin{aligned}
    \Delta s'^{\textrm{can}}_t=&s'^{\textrm{can}}_t-s^{\textrm{can}}_{i(t)}\\  
    =&\ln\frac{(n+1+w^*_i+\Delta w^*_i)^{n+1+w^*_i+\Delta w^*_i}}{{(w^*_i+\Delta w^*_i)}^{w^*_i+\Delta w^*_i}(n+1)^{n+1}}\\
    &-\ln\frac{(n+w^*_i)^{n+w^*_i}}{{w^*_i}^{w^*_i}n^n}
    \end{aligned}
\end{equation}
The value of it is decided by the old total weights of this node $w^*_i$ and the increase weights $\Delta w^*_i$ casued by the new node's connection. 

For the nodes in the set $\mathcal{O}$, the growth of the local entropy is equal to 
\begin{equation}
    \begin{aligned}
    \Delta s'^{\textrm{can}}_h=&s'^{\textrm{can}}_h-s^{\textrm{can}}_{i(h)}\\
    =&\ln\frac{(n+1+w^*_i)^{n+1+w^*_i}n^n}{(n+1)^{n+1}(n+w^*_i)^{n+w^*_i}}.
    \end{aligned}
\end{equation}
The same as we already mentioned before, the value of the local entropy growth for the nodes in the set $\mathcal{O}$ is constrained by the network's increasing size $n+1$ and the total weights $w^*_i$. 

\begin{figure}[htbp]
    \centering
        \includegraphics[width=4.4cm]{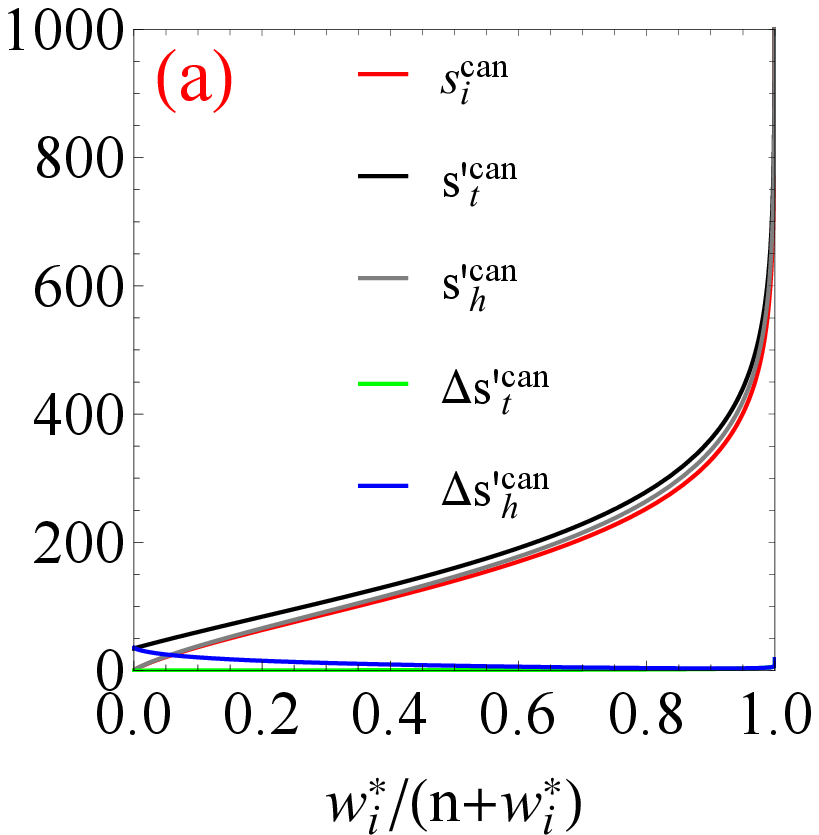}
        \includegraphics[width=4.1cm]{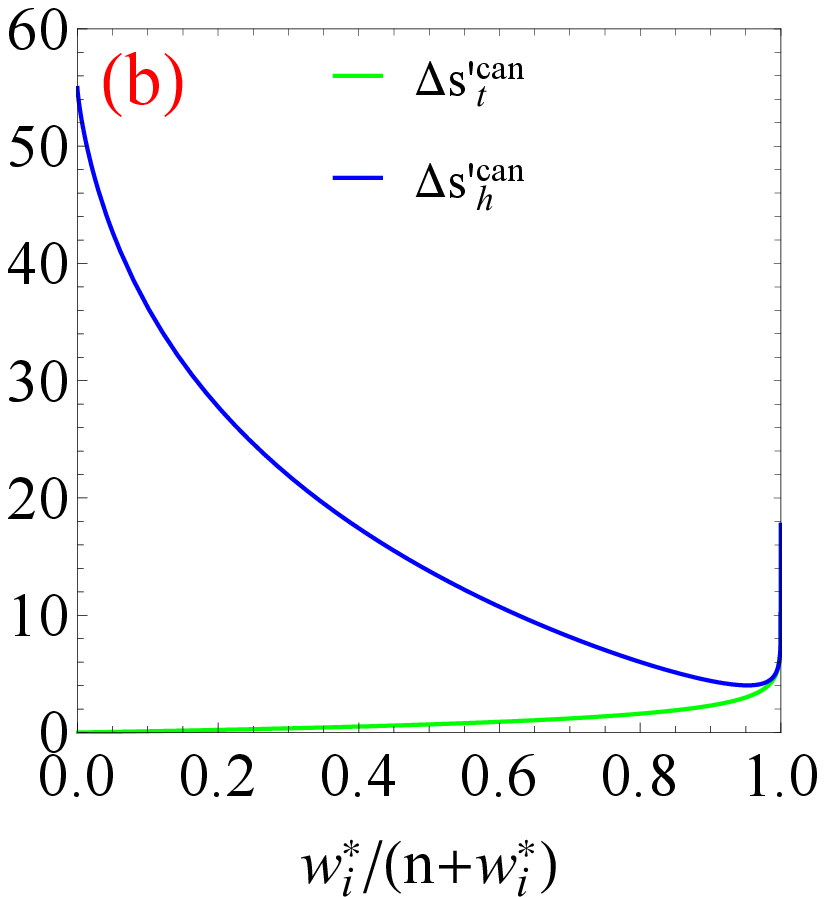}
        \caption{The asymptotic behavior of local entropy in the weighted canonical network ensemble (The original network has 100 nodes). The increase of the total weights of the node shown in these figures is fixed as $20$. Subfigure (a) shows the asymptotic behavior of the local entropy in the weighted canonical network ensemble. We can find that the local entropy of node $i$ with $n$ nodes in it and $n+1$ nodes in it does not have a big difference (both in the set $\mathcal{E}$ and $\mathcal{O}$), especially when the total weights $w^*_i$ of each specific node are close to $\infty$ ($w^*_i/(n+w^*_i)\rightarrow 1$).
        The growth of the node's local entropy in the first way of the node's add has two different paths. The value of $\Delta s'^{\textrm{can}}_t$ is decreasing following the increase of $w^*_i$. It means comparing the increase of the network's size, the influence of the increase of total weights from the newly added nodes is decreased. However, for the node in the set $\mathcal{O}$, the influence of the increasing weights from the newly added node plays an important role.}
        \label{First_weighted_local_entropy_can}
\end{figure}

The details of the asymptotic behavior of the nodes' local entropy are shown in Fig.~\ref{First_weighted_local_entropy_can}. It also includes the asymptotic behavior of the growth of the local entropy. 

The asymptotic behavior of the local entropy in the weighted network ensemble shows that the former total weights decide the value of each node's local entropy. The two different kinds of local entropy are all subleading by the growth of $w^*_i$. But the growth of total weights not always brings an increase in local entropy and its affiliates parameters. We can also find that the value of $\Delta s'^{\textrm{can}}_t$ even drops in the process of $w^*_i$ increasing.

The local entropy of nodes in the two different sets ($\mathcal{E}$ and $\mathcal{O}$) increase when the new node is added to the ensemble, which means the weighted canonical network ensemble is also nonextensive. 

The second way of the node addition in the weighted canonical network ensemble will not change the total weights of those nodes. Thus, the addition of the new node to the weighted network ensemble makes the constraints $\vec{C}(\mathbf{G})=\{\cdots w^*_i, \cdots, w^*_n, w^*_{n+1}\}$. This change in the network ensemble's size also will cause the change in the average value of the total weights as
\begin{equation}
    \langle\tilde{a}_{ij}\rangle=\frac{e^{-\tilde{\theta}^*_i}}{1-e^{-\tilde{\theta}^*_i}}=\frac{w^*_i}{n+1}.
\end{equation}
Then the value of the maximum likelihood parameter $\tilde{\theta}^*_i$ equals to 
\begin{equation}
    e^{-\tilde{\theta}^*_i}=\frac{w^*_i}{n+1+w^*_i}.
\end{equation}
The maximum likelihood parameter of the new network ensemble with the node's addition under the second way's node addition is decided by the total number of nodes in the network ensemble and each node's old total weights. 

This parameter also decides the local entropy of each node in the new network ensemble. Its value equals to 
\begin{equation}
    \begin{aligned}
    \tilde{s}^{\textrm{can}}_{i}&=(n+1)\times[\tilde{\theta}^*_i\frac{ e^{-\tilde{\theta}^*_i}}{1- e^{-\tilde{\theta}^*_i}}+\ln\frac{1}{1- e^{-\tilde{\theta}^*_i}}]\\
    &=[{w^*_i}\ln\frac{n+1+w^*_i}{w^*_i}+(n+1)\ln\frac{n+1+w^*_i}{n+1}]\\
    &=\ln\frac{(n+1+w^*_i)^{n+1+w^*_i}}{{w^*_i}^{w^*_i}(n+1)^{n+1}}  
    \end{aligned}.
\end{equation}

The local entropy $\tilde{s}^{\textrm{can}}_{i}$ has the same definition of $s'^{\textrm{can}}_h$. The asymptotic behavior of $\tilde{s}^{\textrm{can}}_{i}$ also similar with $s'^{\textrm{can}}_h$, and the growth of this local entropy is the same as $\Delta s'^{\textrm{can}}_h$.
Thus, in the weighted canonical network ensemble, adding new nodes to it will change the local entropy of every old node, and the whole structure is nonextensive. 

\section{Local entropy of microcanonical network ensembles}
The microcanonical ensemble in traditional statistical physics is used to describe the systems with fixed total energy. In the network ensemble, when the degree sequences $\{k^*_i\}$ for each network are fixed, i.e., when each node's degree in different networks that belong to a specific ensemble is the same, i.e., the degree sequences $\{k^*_i\}$ is a 'hard' constraint. Therefore, networks in this set need to be described by the microcanonical network ensemble. 

Because the constraint of each network in the microcanonical network ensemble is the same, so the probability distribution of those networks in the microcanonical ensemble is uniform, i.e., the probability of each network is the same, and the value of this probability is decided by the total number of networks under this 'hard' degree sequence.

When the total number of networks in the microcanonical ensemble $\mathcal{G}_{\textrm{mic}}(n,\{{k^*_i}\})$ (with total nodes' number $n$ and degree sequences $\{k^*_i\}$) is represented by $\Omega_{\textrm{mic}}$, the probability of each network $\mathbf{G}$ in the microcanonical ensemble should equal to
\begin{equation}\label{P_mic_degree}
    P_{\textrm{mic}}(\mathbf{G})=\frac{1}{\Omega_{\textrm{mic}}}.
\end{equation}
Then the Shannon entropy of the microcanonical network ensemble $\mathcal{G}_{\textrm{mic}}(n,\{{k_i}\})$ based on the probability distribution $P_{\textrm{mic}}(\mathbf{G})$ or the total number of the networks in this ensemble is
\begin{equation}\label{S_mic_degree}
    S_{\textrm{mic}}=\ln\Omega_{\textrm{mic}}
\end{equation}
It means the number of configurations of networks decides the complexity of the microcanonical network ensemble under 'hard' constraints. 
However, calculating $\Omega_{\textrm{mic}}$ is difficult, and simultaneously, to build the definition of local entropy, we need to localize the calculation of $\Omega_{\textrm{mic}}$. 

\subsection{Local entropy in binary microcanonical network ensemble}
The value of $\Omega_{\textrm{mic}}$ in a network ensemble with $n$ nodes and degree sequence $\{k_i\}$ is decided by the value of units in it. When the existing or not of a link between node $i$ and node $j$ in the network $\mathbf{G}$ that belongs to the microcanonical ensemble $\mathcal{G}_{\textrm{mic}}(n,\{{k_i}\})$ is represented by the value of $1$ or $0$ ($g_{ij}=1$ or $g_{ij}=0$), these networks are the binary network and the number of the configurations $\Omega_{\textrm{mic}}$ of this network ensemble with $\{k_i^*\}$ is equal to 
\begin{equation}\label{The_connection}
   \Omega_{\textrm{mic}}=\prod_{i=1}^n\binom{n}{k_i^*},
\end{equation}
where $\{k_i^*\}$ represents the degree sequences $\{k_i\}$ with a specific value, and $n$ is the total number of nodes in the network $\mathbf{G}$. Eq.\eqref{The_connection} shows that when the constraints of the network ensemble are localized, the calculation of $\Omega_{\textrm{mic}}$ is also localized, and based on this localization we can give the definition of local entropy in the binary microcanonical network ensemble. Before that, we can give the Shannon entropy of the binary microcanonical network ensemble as
\begin{equation}
    S_{\textrm{mic}}=\ln\Omega_{\textrm{mic}}=\sum_{i=1}^n\ln\binom{n}{k^*_i}.
\end{equation}
The Shannon entropy equals the sum of $\ln\binom{k^*_i}{n}$, which is based on the possible distribution of the $k^*_i$ links that are connected with node $i$ among other $n$ nodes (include node $i$ itself). This distribution of the node $i$'s links can be used to calculate the local entropy of node $i$, which means the definition of the node $i$'s local entropy is
\begin{equation}
    s^{\textrm{mic}}_i=\ln\binom{n}{k^*_i}.
\end{equation}
This result proved that in the binary microcanonical network ensemble, the entropy of the whole network ensemble is equal to the sum of all the node's local entropy. The local entropy is based on localized constraints, so their local entropy is also independent when the constraints are independent. 

\subsubsection{The nonextensivity in the binary microcanonical network ensemble. }
The extensivity is the macroscopic property increasing when the system's size grows. Thus, to check if the increase of the system's size will bring a linear increase in the system's macroscopic property, we need to add a new node in the network ensemble with degree $k^*_{n+1}$.

The new node's addition to the microcanonical network ensemble still has two different ways, as shown in Fig.~\ref{Two_ways_node_add}. The first way still aims to keep the original network's structure fixed and add the new links and nodes to the old network's structure. Thus, when the newly added node has degree $k^*_{n+1}$, there will be $k^*_{n+1}$ nodes in the new network ensemble with degree $k^*_i+1$ (the symbol $k^*_i$ represents the old degree of those nodes). When the new node is added to the binary microcanonical network ensemble, the nodes can be divided into two sets. Set $\mathcal{E}$, which includes all the nodes that are connected with the new node; set $\mathcal{O}$, which include all the node whose degree is fixed (this set also has the new node $n+1$).

For the nodes in the set $\mathcal{E}$ with degree $k^*_i+1$, the local entropy of them equals to 
\begin{equation}
    s'^{\textrm{mic}}_{t}=\ln\binom{n+1}{k^*_i+1},
\end{equation}
where $t$ represents the new label of the node $i$ when it is included in the set $\mathcal{E}$. There are $k^*_{n+1}$ nodes in this set, so the Shannon entropy of nodes in the set $\mathcal{E}$ is 
\begin{equation}
    S'^{(\mathcal{E})}_{\textrm{mic}}=\sum_{t=1}^{k^*_{n+1}}s'^{\textrm{mic}}_{t}.
\end{equation}

For the node in the set $\mathcal{O}$, which keeps the old degree as the new degree in the new network ensemble, the value of their local entropy equals to 
\begin{equation}
    s'^{\textrm{mic}}_{h}=\ln\binom{n+1}{k^*_i},
\end{equation}
where $h$ represents the new label of node $i$ in the set $\mathcal{O}$. 

Obviously, there is $n-k^*_{n+1}+1$ nodes in the set $\mathcal{O}$ and their Shannon entropy equal to the sum of all the $n-k^*_{n+1}+1$ nodes' local entropy as 
\begin{equation}
    S'^{(\mathcal{O})}_{\textrm{mic}}=\sum_{h=1}^{n-k^*_i+1}s'^{\textrm{mic}}_{h}.
\end{equation}
In the calculation, the local entropy of the new node $n+1$ also belongs to the set $\mathcal{O}$, and its local entropy equal to 
\begin{equation}
    s'^{\textrm{mic}}_{n+1}=\ln\binom{n+1}{k^*_{n+1}}.
\end{equation}

The local entropy growth for the nodes in the two sets is different. For instance, the local entropy for the node in the set $\mathcal{E}$, the increase of their local entropy $\Delta s'^{\textrm{mic}}_h$ is 
\begin{equation}
    \begin{aligned}
        \Delta s'^{\textrm{mic}}_t&=s'^{\textrm{mic}}_t-s^{\textrm{mic}}_i\\
        &=\ln\binom{n+1}{k^*_i+1}-\ln\binom{n}{k^*_i}\\
        &=\ln\frac{n+1}{k^*_i+1}
    \end{aligned}.
\end{equation}

For the nodes in the set $\mathcal{O}$, the growth of their local entropy is also constrained by the network's size and the old degree as 
\begin{equation}
    \begin{aligned}
        \Delta s'^{\textrm{mic}}_h&=s'^{\textrm{mic}}_h-s^{\textrm{mic}}_i\\
        &=\ln\binom{n+1}{k^*_i}-\ln\binom{n}{k^*_i}\\
        &=\ln\frac{n+1}{n-k^*_i+1}
    \end{aligned}.
\end{equation}

We can find that the value of the Shannon entropy increase when the new node adds to the network ensemble with the first way, and the increasing value of it equals to 
\begin{equation}
    \Delta S'_{\textrm{mic}}=\sum_{t=1}^{k^*_{n+1}}\Delta s'^{\textrm{mic}}_t+\sum_{h=1}^{n-k^*_{n+1}+1}\Delta s'^{\textrm{mic}}_h.
\end{equation}

\begin{figure}[htbp]
    \centering
        \includegraphics[width=4.3cm]{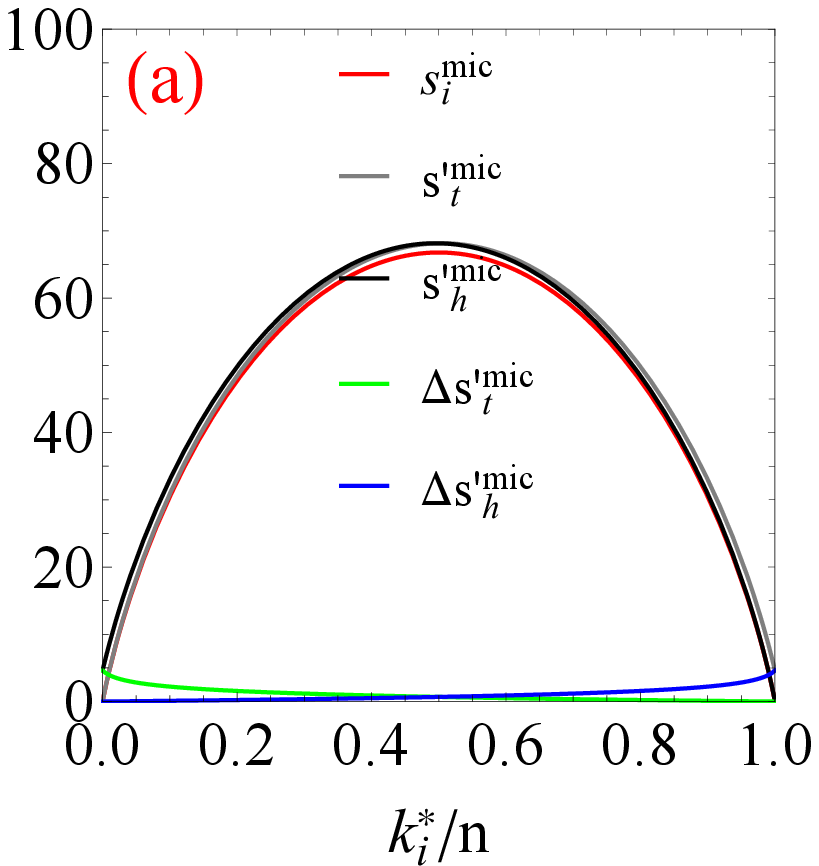}
        \includegraphics[width=4.15cm]{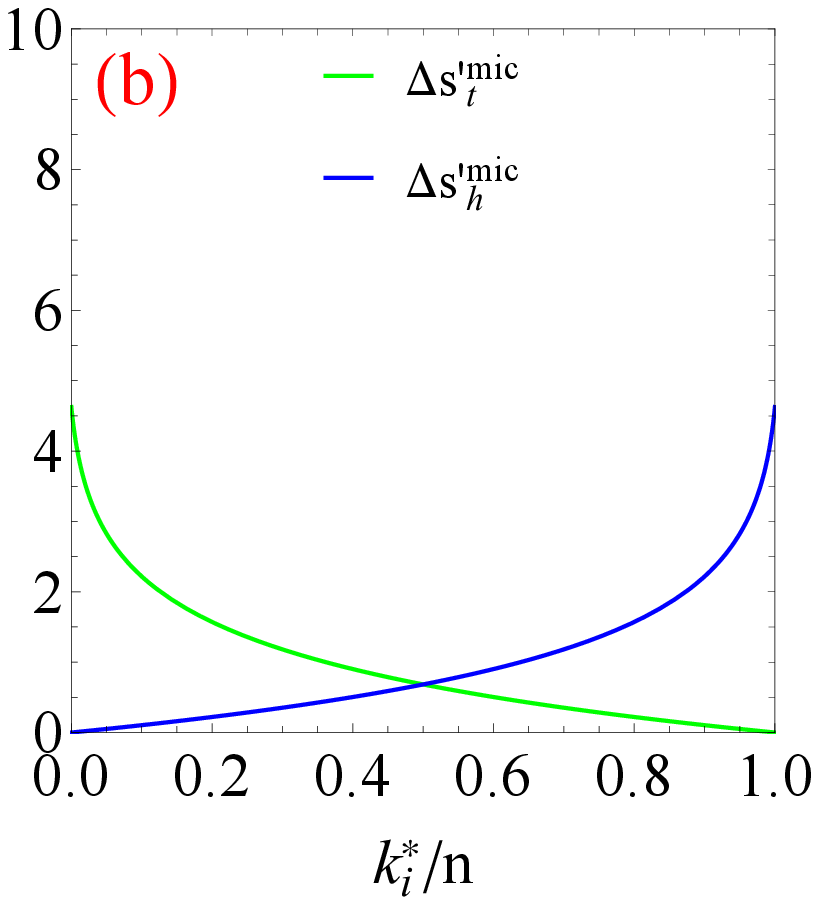}
        \caption{The asymptotic behavior of local entropy in the binary microcanonical network ensemble (The original network still has 100 nodes). Subfigure (a) shows the asymptotic behavior of the local entropy in the binary microcanonical network ensemble. We can find that the local entropy of node $i$ with $n$ nodes in it and $n+1$ nodes in it does not have a big difference, and they are symmetry with the ratio between $k^*_i$ and $n$. when the value of ${k^*_i}/{n}$ is equal to $0.5$, the local entropies have the biggest value (both for the $s^{\textrm{mic}}_{i}$, ${s'}^{\textrm{mic}}_{t}$ and ${s'}^{\textrm{mic}}_{h}$). The growth of the node's local entropy in the first way of node's add has two different paths, the bigger the original degree of the node, the bigger the value of the local entropy growth when the nodes are node include in the set $\mathcal{O}$. However, in the set of $\mathcal{E}$, the bigger the original degree of the node, the smaller the increase of local entropy.}
        \label{First_Binary_local_entropy_mic}
\end{figure}

The asymptotic behavior of local entropy in the two sets of the binary microcanonical network ensemble are shown in Fig.~\ref{First_Binary_local_entropy_mic}

The asymptotic behavior of the local entropy in the microcanonical ensemble is similar to that in the canonical ensemble.

In the second way of the node's addition, the degree of all the existing nodes in the network ensemble will keep fixed, and the constraints in the network ensemble are fixed from their numeric value. Only the network size will increase $1$ after adding new $k^*_{n+1}$ subs to the network ensemble.  

As there are $n+1$ nodes in the network ensemble, the calculation of Shannon entropy will become the sum of all the $n+1$ nodes' local configurations. The value of the Shannon entropy $\tilde{S}_{\textrm{mic}}$ of the new network ensemble with $n+1$ nodes should equal to
\begin{equation}
    \tilde{S}_{\textrm{mic}}=\sum_{i=1}^{n+1}\ln\binom{n+1}{k^*_i}.
\end{equation}
Then the local entropy of node $n+1$ is based on the possible configurations of the $k^*_{n+1}$ links among all the $n+1$ nodes (still include node $n+1$ it'self), and the value of it is
\begin{equation}
    \tilde{s}_{n+1}^{\textrm{mic}}=\ln\binom{n+1}{k^*_{n+1}}.
\end{equation}
Simultaneously, we can find the local entropy of node $i$ also equals to the possible configurations of its $k^*_i$ links among $n+1$ nodes, and it is equal to
\begin{equation}
    \tilde{s}^{\textrm{mic}}_{i}=\ln\binom{n+1}{k^*_i}.
\end{equation}

There is a biased assumption in the new network ensemble that the new add nodes will connect with the other $k^*_{n+1}$ nodes in the existing $n$ nodes, but the degree of those $k^*_{n+1}$ nodes does not change. It means the first $n$ units in the degree sequence of the new network ensemble are the same as the original one, but the structure of the networks will change. This difference in the network structure can be quantified by the difference in the Shannon entropy of the two network ensembles as
\begin{equation}
    \begin{aligned}
        \Delta \tilde{S}_{\textrm{mic}}&=\tilde{S}_{\textrm{mic}}-S_{\textrm{mic}}\\
        &=\ln\tilde{\Omega}_{\textrm{mic}}-\ln\Omega_{\textrm{mic}}\\
        &=\ln{\prod_{i=1}^{n+1}\binom{n+1}{k_i^*}}-\ln{\prod_{i=1}^n\binom{n}{k_i^*}}\\
        &=\ln\binom{n+1}{k^*_{n+1}}+\sum_{i=1}^n\ln\frac{n+1}{n-k^*_i+1},
    \end{aligned}
\end{equation}
where $\tilde{\Omega}_{\textrm{mic}}$ represents the total number of configurations for the network ensemble with degree sequence $\{k^*_i\}$ (with n+1 nodes). As the $\ln\binom{n+1}{k^*_{n+1}}$ is the local entropy of node $n+1$ in our setting, thus, the increasing of the structural complexity in the original network is $\sum_{i=1}^n\ln\frac{n+1}{n-k^*_i+1}$.

The extra part in the $\Delta S_{\textrm{mic}}$ is a sum of $n$ different part. It may connect with the increase of each node's local entropy. To illustrate the relationship between the increase of the network ensemble's structural complexity and the growth of each node's local entropy, we fid the the growth of each node's local entropy as 
\begin{equation}
    \begin{aligned}
    \Delta \tilde{s}^{\textrm{mic}}_i&=\tilde{s}^{\textrm{mic}}_i-s^{\textrm{mic}}_i\\
    &=\ln\binom{n+1}{k^*_i}-\ln\binom{n}{k^*_i}\\
    &=\ln\frac{n+1}{n-k^*_i+1} 
    \end{aligned}.
\end{equation} 
Thus, the extra part in $\Delta S_{\textrm{mic}}$ is the sum of the increase of the $n$ nodes' local entropy. The value of $\Delta S_{\textrm{mic}}$ is equal to the sum of the structural complexity change in the original $n$ nodes and the local entropy of the added node $n+1$,
\begin{equation}
    \Delta \tilde{S}_{\textrm{mic}}=\tilde{s}_{n+1}^{\textrm{mic}}+\sum_{i=1}^n\Delta \tilde{s}^{\textrm{mic}}_i.
\end{equation}

This result proves that the newly added node (node with label $n+1$) and its connections with other nodes in the network bring the growth of the structural complexity of the network ensemble. This growth is not extensive, as the newly added node not only brings its local entropy to the network ensemble but also brings an increase in all the nodes' local entropy, i.e., the binary microcanonical network ensemble is nonextensive. 

\subsection{Local entropy in weighted microcanonical network ensemble}
In network science, when the links between different pairs of nodes have different weights, the configurations of this network ensemble with degree sequences $\vec{C}^*=\{w^*_i\}$ are different from the binary one. The set of these networks is the weighted network ensemble. When the degree sequences of each weighted network in the networks ensemble are strictly fixed the same, it is a weighted microcanonical network ensemble. 

The weighted network has a different configuration compared with the binary network ensemble. Thus, the number of configurations $\Omega_{\textrm{mic}}$ for the weighted network is different from the binary one. And it is difficult to calculate. However, we can still use the relationship between the canonical ensemble and the conjugate microcanonical ensemble to estimate the probability of each state of the network ensemble because the microcanonical ensemble is a subset of the conjugate canonical ensemble when the constraints of those networks have constraints $\vec{C}^*$~\cite{squartini2017reconnecting,zhang2022strong}.

For the weighted network ensemble with total weights' sequence $\{w^*_i\}$, the number of networks that satisfies the constraints $\vec{C}(\mathbf{G})=\vec{C}^*$ in the canonical ensemble can be calculated by the Delta-Dirac function and canonical ensemble when the maximum likelihood parameter $\vec{\theta^*}$ is extended to complex number $\vec{\theta^*}+i\vec{\psi}$,
\begin{equation}
    \begin{aligned}
    \Omega_{\textrm{can}}&=\sum_{\mathbf{G}\in\mathcal{G}_{\textrm{can}}}\int^{+\vec{\pi}}_{-\vec{\pi}}\frac{\,d\vec{\psi} }{(2\pi)^n}e^{i\vec{\psi}[\vec{C}^*-\vec{C}(\mathbf{G})]}\\ 
    &=\int^{-\vec{\pi}}_{-\vec{\pi}}\frac{\,d\vec{\psi} }{(2\pi)^n}P^{-1}_{\textrm{can}}(\mathbf{G}^*|\vec{\theta^*}+i\vec{\psi})\\
    &=\prod_{i=1}^n\int_{-\pi}^{+\pi}\frac{\,d{\psi_i}}{2\pi}e^{-(\theta^*_i+i\psi_i)-w^*_i}[1-e^{-(\theta^*_i+i\psi_i)}]^{-n}
    \end{aligned}
\end{equation} 
This results still far away with the specific number of the configurations, so we change the variable $y=e^{-(\theta^*_i+i\psi_i)}$, $\,d\psi=i\,d y/y$ and rearrange the integral as 
\begin{equation}
    \begin{aligned}
        \Omega_{\textrm{can}}&=\frac{i}{2\pi}\prod_{i=1}^n\int_{e^{-(\theta^*_i+i\pi)}}^{e^{-(\theta^*_i-i\pi)}}y^{-(w^*_i+1)}{(\frac{1}{1-y})^n}\,d y\\
        &=\frac{i}{2\pi}\prod_{i=1}^n\int_{e^{-(\theta^*_i+i\pi)}}^{e^{-(\theta^*_i-i\pi)}}y^{-(w^*_i+1)}{(1+\frac{y}{1-y})^n}\,d y\\
    \end{aligned}.
\end{equation}
Then we perform another change of variable $z=y/(1-y)$, $\,dy=\,dz/(z+1)^2$, we can have the new form of the $\Omega_{\textrm{mic}}$ of the weighted network from the binomial formula as 
\begin{equation}
    \begin{aligned}
        \Omega_{\textrm{can}}&=\prod_{i=1}^n\frac{i}{2\pi}\int_{z^+}^{z^-}(\frac{z}{z+1})^{-(w^*_i+1)}\frac{(1+z)^n}{(z+1)^2}\,d z\\
        &=\prod_{i=1}^n\frac{i}{2\pi}[\sum_{k=0}^{n}\binom{n}{k}\int_{z^+}^{z^-}(\frac{z}{z+1})^{-(w^*_i+1)}\frac{z^k}{(z+1)^2}\,d z]\\
        &=\prod_{i=1}^n\frac{i}{2\pi}[\sum_{k=0}^{n}\binom{n}{k}\int_{z^+}^{z^-}{z^{k-(w^*_i+1)}}{(z+1)^{w^*_i-1}}\,d z]\\
        &=\prod_{i=1}^n\frac{i}{2\pi}[\sum_{k=0}^{n}\binom{n}{k}\sum_{h=0}^{k^*_i-1}\binom{w^*_i-1}{h}\int_{z^+}^{z^-}{z^{k-(w^*_i+1)+h}}\,d z]\\        
    \end{aligned}
\end{equation}
The $z^-$ and $z^+$ are defined as 
\begin{equation}
    z^{\pm}=\frac{e^{-\theta^*_i\pm{i\pi}}}{1-e^{-\theta^*_i\pm{i\pi}}}.
\end{equation}
To get the final result, we still need to use the residue theorem that the only non-zero integral is when the value of $h=w^*_i-k$, and the value of the integral is $-2\pi i$, we can have the value of $\Omega_{\textrm{mic}}$ as 
\begin{equation}\label{Omega_W+f}
    \begin{aligned}
    \Omega_{\textrm{mic}}&=\prod_{i=1}^n[\frac{i}{2\pi}\sum_{k=0}^{n}\binom{n}{k}\binom{w^*_i-1}{w^*_i-k}(-2\pi i)]\\
    &=\prod_{i=1}^n[\sum_{k=0}^{n}\binom{n}{k}\binom{w^*_i-1}{w^*_i-k}]\\
    &=\prod_{i=1}^n\binom{n+w^*_i-1}{w^*_i}.\\
    \end{aligned}
\end{equation}
Then the Shannon entropy of the weighted network ensemble with degree sequence $\{w^*_i\}$ can be calculated exactly as 
\begin{equation}
    S_{\textrm{mic}}=\ln\Omega_{\textrm{mic}}=\sum_{i=1}^n\ln\binom{n+w^*_i-1}{w^*_i}.
\end{equation}
This result also clearly shows that the local entropy of each node in the weighted network ensemble is constrained by the number of nodes that can be connected and the total strength of weights that can be allocated to each link as
\begin{equation}
    s_{i}^{\textrm{mic}}=\ln\binom{n+w^*_i-1}{w^*_i}.
\end{equation}
It means the local entropy of each node is independent of each other.

\subsubsection{The nonextensivity in the weighted microcanonical network ensemble}
As we already show in the binary microcanonical ensemble, the number of possible configurations in the microcanonical network ensemble is directly decided by its constraints. When we add new nodes to each network ensemble, the number of constraints $\vec{C}(\mathbf{G})$ will increase. And simultaneously, the value of each unit's constraints may also change (the first way of nodes' addition). Or the value of each unit's constraint is fixed (The second way of nodes' addition). All the two different ways of nodes' addition still keep the same calculation form of $\Omega'_{\textrm{mic}}$, but with different specific values.  

Thus, when the new add total strength of the node $n+1$ is $w^*_{n+1}$, the change of the partition function and probability of each network in the new network ensemble will affect the calculation of the $\Omega_{\textrm{mic}}$, we can still follow the existed form of the calculation. 

When the nodes' addition is based on the first way, the nodes in it still can be divided into two sets, set $\mathcal{E}$ and $\mathcal{O}$ (The definitions are the same as the one in the binary microcanonical ensemble.). We assume there are $l$ nodes that will change their total weights (in the set $\mathcal{E})$, and the other $n+1-l$ nodes will keep their total weights (set $\mathcal{O}$). If we set the new total weights of the links that connect with different nodes as ${w'}_{i'}$, where ${i'}\in[1,l]$. Then we can have the value of the $\Omega'_{\textrm{mic}}$ as 
\begin{equation}
    \Omega'_{\textrm{mic}}=\prod_{i'=1}^l\binom{n+{w'}_{i'}}{{w'}_{i'}}\prod_{\tilde{i}=1}^{n+1-l}\binom{n+w^*_{\tilde{i}}}{w^*_{\tilde{i}}},
\end{equation}
where $w^*_{\tilde{i}}$ represents the total weights of nodes ${\tilde{i}}$ that belonged to the set $\mathcal{O}$. The total number of nodes in the network ensemble is $n+1$, so $n$ in Eq.~\eqref{Omega_W+f} will be replaced as $n+1$ and we will get the new calculation of $\Omega'_{\textrm{mic}}$ above. 

Then we can find the new local entropy of nodes in the set $\mathcal{E}$ of the weighted microcanonical network ensemble is 
\begin{equation}
    {s'}^{\textrm{mic}}_{i'}=\ln\binom{n+{w'}_{i'}}{{w'}_{i'}}=\ln\binom{n+w^*_i+\Delta w^*_i}{w^*_i+\Delta w^*_i}.
\end{equation}
For the nodes in the set $\mathcal{O}$, the local entropy of them is 
\begin{equation}
    \tilde{s}^{\textrm{mic}}_{\tilde{i}}=\ln\binom{n+{w'}_{\tilde{i}}}{{w'}_{\tilde{i}}}=\ln\binom{n+w^*_i}{w^*_i}.
\end{equation}
The total weights of the node ${i'}$ in the set $\mathcal{E}$ are equal to ${w'}_{i'}={w}^*_{i}+\Delta{w}^*_{i}$ (here $i$ represents the label of nodes ${i'}$ mapping as its original label in the network ensemble). Thus, the growth of the local entropy of the nodes in the set $\mathcal{E}$ equals 
\begin{equation}
    \begin{aligned}
    \Delta {s'}^{\textrm{mic}}_{i'}&={s'}^{\textrm{mic}}_{i'}-{s}^{\textrm{mic}}_{i}\\
    &=\ln\binom{n+w^*_i+\Delta w^*_i}{w^*_i+\Delta w^*_i}-\ln\binom{n+w^*_i-1}{w^*_i}\\
    &=\ln\frac{\prod_{\tilde{a}=1}^{\Delta w^*_i+1}(n+w^*_i-1+\tilde{a})}{n\times\prod_{\tilde{b}=1}^{\Delta w^*_i}(w^*_i+\tilde{b})}.
    \end{aligned}
\end{equation}

For the nodes in the set $\mathcal{O}$, the growth of their local entropy is decided by the value of ${w'}_{i'}$, which is equal to ${w'}_{i'}=w^*_i$.
The value of this local entropy growth is 
\begin{equation}
    \begin{aligned}
    \Delta \tilde{s}^{\textrm{mic}}_{\tilde{i}}&=\tilde{s}^{\textrm{mic}}_{\tilde{i}}-{s}^{\textrm{mic}}_{i}\\
    &=\ln\binom{n+w^*_i}{w^*_i}-\ln\binom{n+w^*_i-1}{w^*_i}\\
    &=\ln\frac{n+w^*_i}{n}      
    \end{aligned}
\end{equation}

\begin{figure}[htbp]
    \centering
        \includegraphics[width=4.35cm]{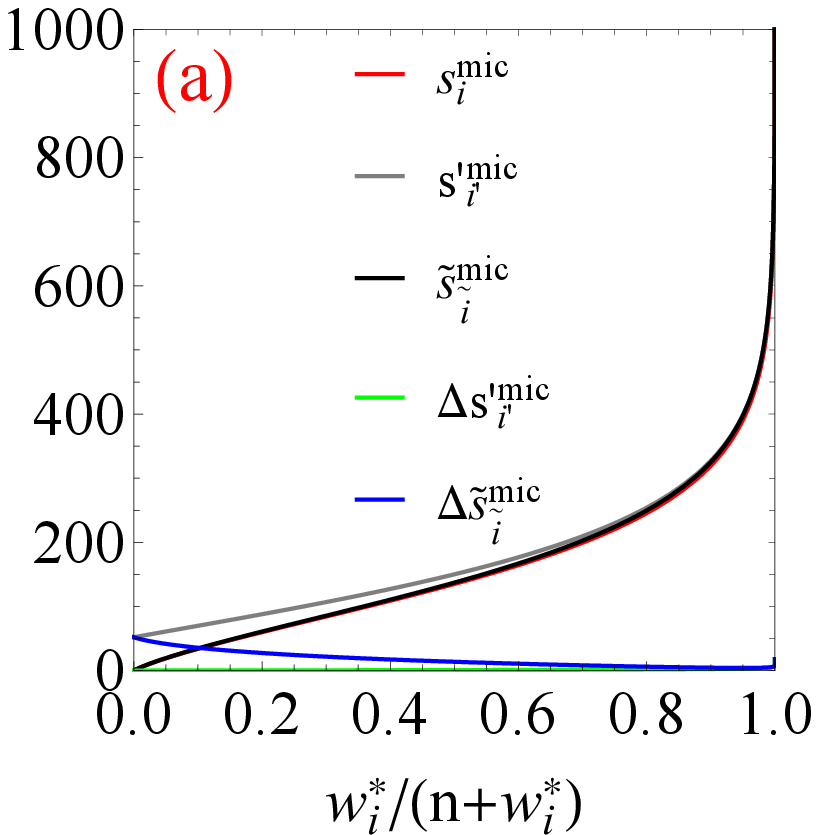}
        \includegraphics[width=4.1cm]{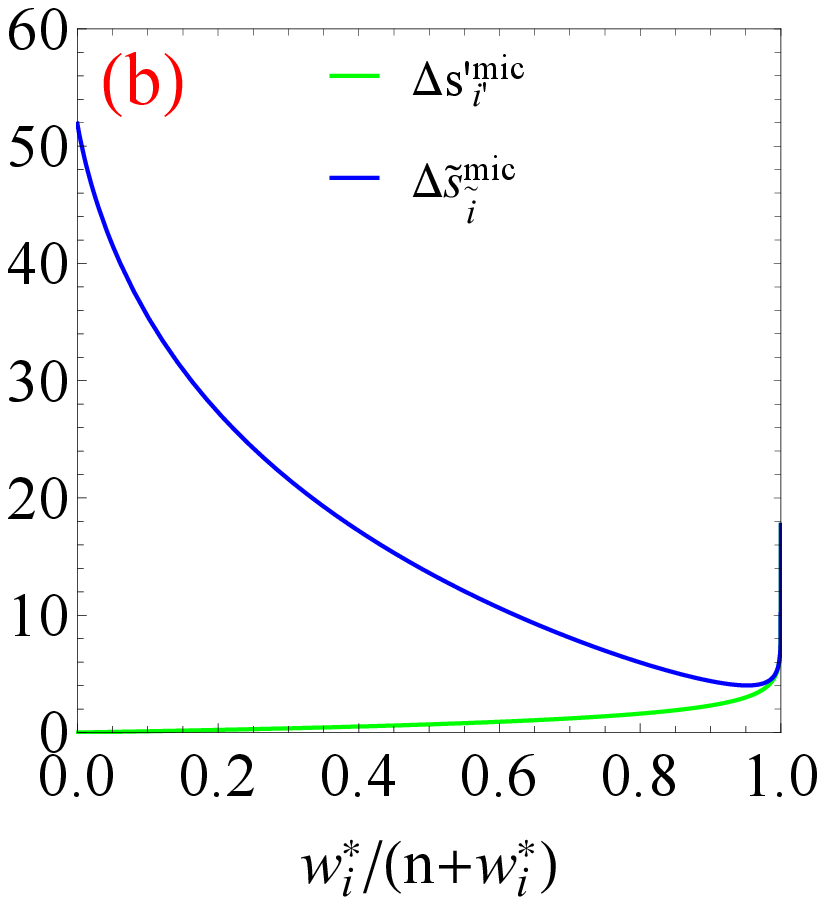}
        \caption{The asymptotic behavior of local entropy in the weighted microcanonical network ensemble (The original network still has 100 nodes). Subfigure (a) shows the asymptotic behavior of the local entropy in the weighted microcanonical network ensemble. We can find that the local entropy of node $i$ with $n$ nodes in it and $n+1$ nodes in it does not have a big difference, and their value is increased following the growth of the total weights. When the value of ${w^*_i}/{n+w^*_i}$ is equal to $1$, the local entropies have the biggest value (both for the $s^{\textrm{mic}}_{i}$, ${s'}^{\textrm{mic}}_{i'}$ and $\tilde{s}^{\textrm{mic}}_{\tilde{i}}$). The growth of the node's local entropy in the first way of the node's adding has two different paths, the bigger the original degree of the node, the bigger the value of the local entropy growth when the nodes are node included in the set $\mathcal{O}$. However, in the set of $\mathcal{E}$, the bigger the original degree of the node, the smaller the increase of local entropy.}
        \label{First_weighted_local_entropy_mic}
\end{figure}

The asymptotic behavior of the local entropy is decided by the value of $w^*_i$ and the growth of total weights $\Delta w^*_i$. The details of their asymptotic behavior are shown in Fig.~\ref{First_weighted_local_entropy_mic}.

The results in Fig.~\ref{First_weighted_local_entropy_mic} show that the local entropy of the nodes in the weighted microcanonical network ensemble has the same asymptotic behavior, which means from the scale of each node, there is ensemble equivalence. 

When the node's addition to the weighted microcanonical network ensemble uses the second-way shown in Fig.~\ref{Two_ways_node_add} the number of the configurations in the new microcanonical ensemble is
\begin{equation}
    \begin{aligned}
        \Omega'_{\textrm{mic}}&=\prod_{i=1}^{n+1}\frac{i}{2\pi}\int_{z^+}^{z^-}(\frac{z}{z+1})^{-(w^*_i+1)}\frac{(1+z)^{n+1}}{(z+1)^2}\,d z\\
        &=\prod_{i=1}^{n+1}\frac{i}{2\pi}[\sum_{k=0}^{n+1}\binom{n+1}{k}\int_{z^+}^{z^-}{z^{k-(w^*_i+1)}}{(z+1)^{(w^*_i-1)}}\,d z]\\
        &=\prod_{i=1}^{n+1}\frac{i}{2\pi}[\sum_{k=0}^{n+1}\binom{n+1}{k}\sum_{h=0}^{w^*_i-1}\binom{w^*_i-1}{h}\int_{z^+}^{z^-}{z^{k-(w^*_i+1)+h}}\,d z]\\
        &=\prod_{i=1}^{n+1}[\sum_{k=0}^{n}\binom{n+1}{k}\binom{w^*_i-1}{w^*_i-k}]\\
        &=\prod_{i=1}^{n+1}\binom{n+w^*_i}{w^*_i}.
    \end{aligned}
\end{equation}
Then the Shannon entropy and local entropy of the new weighted canonical network ensemble equal to 
\begin{equation}
    S'_{\textrm{mic}}=\sum_{i=1}^{n+1}\ln\binom{n+w^*_i}{w^*_i}, s'^{\textrm{mic}}_i=\ln\binom{n+w^*_i}{w^*_i}.
\end{equation}
The newly added node with label $n+1$ also brings the increase of each node's local entropy in the weighted canonical network ensemble. The growth of the local entropy of node $i$ is equal to 
\begin{equation}
    \begin{aligned}
    \Delta s^{\textrm{mic}}_i&=\ln\binom{n+w^*_i}{w^*_i}-\ln\binom{n+w^*_i-1}{w^*_i}\\
    &=\ln\frac{n+w^*_i}{n},  
    \end{aligned}
\end{equation}
which is the same as the local entropy of nodes in the set $\mathcal {} $, when the nodes' addition uses the first way. 

This result also proves that the weighted canonical network ensemble is nonextensive, and the nonextensivity comes from the heterogeneous interaction among nodes in the network.

\section{Conclusions}
Extensivity in physics is the extensive properties of matter that depend on system size. When the size of the system increases, the value of the extensive properties also grows. One of the typical extensive properties of the system is its entropy. When the size of the system grows, its entropy also increases. Thus, checking how the entropy of the systems changes when the size of the system grows is an effective way to understand its extensivity. 

As we already mentioned, the Shannon entropy of the network ensembles should increase following the growth of the network's size, which is the number of nodes in the network ensembles. However, this is not enough. The research on extensivity still needs to focus on how the entropy changes when a specific node is added to the network ensembles. 
Simultaneously, the extensivity represented by the growth of entropy in the systems has a distinct phenomenon that the entropy of the old system does not change with the increase of new units to the system, but in the network ensembles, this phenomenon can not maintains. The newly added node in the network ensemble not only brings new configurations for the network ensemble but also changes the local structure of those existing nodes. It means from the global view, every newly added node in the network ensemble will bring an increase in the network ensemble's structural complexity. Simultaneously, it may also change the local structural complexity of each node. Thus, we proposed the local entropy of nodes in the network ensembles, and we used the change of the value of each node's local entropy to prove that the network ensemble is nonextensive. 
The local entropy defined in this work is based on the local structure around each node. Thus, the change in the local structural characteristics and the size of the network ensemble will affect the value of each node's local entropy. 

In the canonical network ensemble, each node is independent. The local structural complexity is decided by the degree or total weights (binary or weighted network) for each node and the size of the network ensembles. Thus, when a new node is added to the network ensemble, the degree or the total weights for each node may change. As the newly added node will bring new links to the network ensemble, the change or not of the node's degree or total weights is still decided by way of the nodes' addition. For instance, when the node's addition is based on the first way, the value of the degree or total weights for the nodes directly connected with the new node will change. However, when the node's addition is based on the second way, the degree or total weights for each old node will be fixed the same, but the size of the network ensemble (the number of nodes in it) will grow. Thus, no matter which way to add the nodes to the network ensemble, the network size will change. Simultaneously, the possible configurations of the weights or links' distribution will also change, i.e., the local entropy will grow. 

The same phenomenon also happens in the microcanonical network ensemble. Even though all the possible dependencies are counted in the local entropy's calculation, each node's local entropy is still independent will other nodes. But the growth of the system's size still brings the change of all of the nodes' local entropy, i.e., the change of the local structural complexity in the network ensemble, both in the weighted and binary network ensemble. 

Therefore, in the network ensemble, each node's local entropy is not independent, i.e., adding new nodes to the network will affect each node's local structure. Thus, the network ensembles are not extensive, i.e., nonextensivity always exists in the network ensemble. 

This nonextensivity in the network ensembles shows why the network model is so powerful when it is used to describe those complex phenomena. As we know in statistical physics, the general model used to simulate those systems is the Ising model, where each unit has 2 degrees of freedom (when the possible degree of freedom of each unit increases from 2 to all the natural numbers, we will have the weighted Ising model, which is called the Potts model). The interaction among their different units manipulates the macroscopic properties of these Ising-like modeled systems. 
The Ising model can be treated as a particular case of the binary network ensembles when only two links are allowed to connect with each node, and the maximum number of neighbors is 4. Ananogly, the Potts model also can be treated as a particular case of the weighted network ensemble. Thus, from the different links connect ways, we can find the distinct difference between the traditional Ising-like model and the networks model is their local extensivity. In the Ising-like model, each unit is independent, and the newly added units will not affect the possible number of configurations of the existing units, i.e., the system's growing size will not affect the structural complexity of the existing parts. But in the systems described by networks, the order in the Ising-like system does not exist. Every unit in the networked systems can build connections with other units. Thus, every newly added node will cause an increase in the structural complexity of the whole system. And this structural complexity increase is nonextensive.

\section*{Acknowledgments}
This work was supported by the Scientific research funding of Jiangsu University of Science and Technology (No.1052932204). We thank Prof. dr. Zhihong You from Xiamen University for his valuable suggestions in the beginning of this works. We would also like to express our great appreciation to Prof. dr. Diego Garlaschelli from Leiden University for his constrctive suggestions during this work. 

\bibliography{qi_test_google.bib}

\end{document}